\documentclass[sigconf]{acmart}

\usepackage{xspace}
\usepackage{xcolor}
\usepackage{soul}
\usepackage{hyperref}

\usepackage{algorithm}
\usepackage{algorithmic}
\usepackage{enumitem}
\usepackage{array}
\usepackage{booktabs}
\usepackage{listings}
\usepackage{varwidth}
\usepackage{framed}

\newcommand{\sys}[0]{Orca\xspace}
\newcommand{\sysnospace}[0]{Orca}
\newcommand{\webcanvas}[0]{Web Canvas\xspace}
\newcommand{\pinnedbar}[0]{Pinned Webpage Bar\xspace}
\newcommand{\extraction}[0]{Page Extraction\xspace}
\newcommand{\batchopen}[0]{Batch Open\xspace}
\newcommand{\expansion}[0]{Contextual Expansion\xspace}
\newcommand{\expose}[0]{Expos\'{e}\xspace}

\newcommand{\code}[1]{\texttt{#1}}

\definecolor{shadecolor}{gray}{0.93}
\newenvironment{scenario}{\begin{shaded*}\samepage\itshape}{\end{shaded*}}

\newcommand{\rv}[1]{#1}

\AtBeginDocument{%
  \providecommand\BibTeX{{%
    \normalfont B\kern-0.5em{\scshape i\kern-0.25em b}\kern-0.8em\TeX}}}

\copyrightyear{2026}
\acmYear{2026}
\setcopyright{cc}
\setcctype{by}
\acmConference[CHI '26]{Proceedings of the 2026 CHI Conference on Human Factors in Computing Systems}{April 13--17, 2026}{Barcelona, Spain}
\acmBooktitle{Proceedings of the 2026 CHI Conference on Human Factors in Computing Systems (CHI '26), April 13--17, 2026, Barcelona, Spain}
\acmDOI{10.1145/3772318.3790335}
\acmISBN{979-8-4007-2278-3/2026/04}

\begin{document}

\title[Orca: Browsing at Scale Through User-Driven and AI-Facilitated Orchestration Across Malleable Webpages]{Orca: Browsing at Scale Through User-Driven and\\AI-Facilitated Orchestration Across Malleable Webpages}

\author{Peiling Jiang}
\email{peiling@ucsd.edu}
\orcid{0000-0003-4447-0111}
\affiliation{%
  \institution{Foundation Interface Lab}
  \institution{University of California San Diego}
  \city{La Jolla}
  \state{California}
  \country{USA}
}

\author{Haijun Xia}
\email{haijunxia@ucsd.edu}
\orcid{0000-0002-9425-0881}
\affiliation{%
  \institution{Foundation Interface Lab}
  \institution{University of California San Diego}
  \city{La Jolla}
  \state{California}
  \country{USA}
}

\renewcommand{\shortauthors}{Jiang and Xia}

\begin{abstract}
  Web-based activities span multiple webpages. However, conventional browsers with stacks of tabs cannot support operating and synthesizing large volumes of information across pages. While recent AI systems enable fully automated web browsing and information synthesis, they often diminish user agency and hinder contextual understanding. We explore how AI could instead augment user interactions with content across webpages and mitigate cognitive and manual efforts. Through literature on information tasks and web browsing challenges, and an iterative design process, we present novel interactions with our prototype web browser, Orca. Leveraging AI, Orca supports user-driven exploration, operation, organization, and synthesis of web content at scale. To enable browsing at scale, webpages are treated as malleable materials that humans and AI can collaboratively manipulate and compose into a malleable, dynamic, and browser-level workspace. Our evaluation revealed an increased ``appetite'' for information foraging, enhanced control, and more flexible sensemaking across a broader web information landscape.
\end{abstract}



\begin{teaserfigure}
  \includegraphics[width=\textwidth]{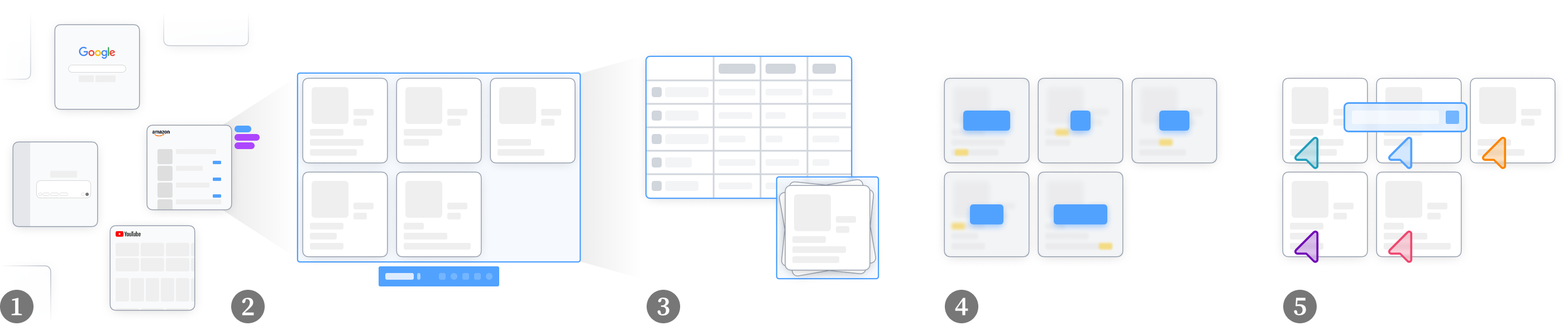}
  \caption{The \sys browser enables \emph{browsing at scale} by supporting viewing (1), navigating (2), organizing (3), extracting (4), operating (5), and synthesizing at scale.}
  \Description{A diagram illustrates the features of the Orca browser in a five-step workflow. Step 1, viewing, shows thumbnails of several different websites. Step 2, navigating, shows a grid of four web pages. Step 3, organizing, depicts the grid view transforming into a table. Step 4, extracting, shows highlighted data elements within the pages in the grid. Step 5, operating and synthesizing, shows colorful cursors pointing at different pages with a command bar, representing an operation across multiple sources.}
  \label{fig:teaser}
\end{teaserfigure}

\maketitle

\section{Introduction}
\label{sec:introduction}

Despite the growing scale and complexity of the web and the information tasks we perform on it, the web browser---the primary interface between users and the web---has seen little fundamental change over the past two decades. The conventional linear stack of tabbed pages has not only fallen short in easing the burden of discovering and reviewing large volumes of information during foraging but also adds considerable overhead in managing and synthesizing information scattered across multiple tabs~\cite{chang2021tab,ma2023browsing}.

Recent advances in Large Language Models (LLMs) have brought transformative changes to how we interact with information on the web. For example, \nobreak{OpenAI}'s Operator integrates a chat interface with a classic browser, which helps translate users' natural language prompts into a sequence of operations on a browser~\cite{openaioperator}. Systems like Deep Research further remove the user from the browsing and synthesis process, delivering a final report~\cite{deepresearch}. While these approaches can reduce the cognitive and manual effort of browsing, managing, and synthesizing content, direct user interaction with web content remains irreplaceable for several reasons.
First, compared to AI-generated summaries, directly engaging with content allows users to consume information in its original context, enabling a more accurate understanding of the problem space~\cite{pirolli1995information}. Second, user-driven browsing lets users stay in control of information sources, steer their evolving exploration directions, preserve a sense of agency, and leave room for serendipity~\cite{foster2003serendipity,teevan2004perfect,suh2023sensecape,pirolli2005sensemaking}.

\rv{Therefore, instead of using AI to take users away from the web, we aim to enhance \emph{web-based information foraging and synthesis}---the most common and demanding uses of browsers~\mbox{\cite{pirolli1995information,sellen2002knowledge,chang2021tab}}---by preserving direct engagement with web content and applying AI to reduce manual and cognitive burden. We explore how web browsers can support \mbox{\textbf{browsing at scale}} for such tasks, allowing users to interact with many webpages in parallel rather than one at a time.}

\rv{Our key approach builds on prior work on canvas-based interactions and malleable software environments by treating webpages as ``malleable materials''---elements that can be viewed in parallel at different scales, summarized, and flexibly extracted---and the browser as a ``malleable space''---a browser-level workspace where these materials can be viewed simultaneously, spatially arranged, grouped, and composed into task-centric structures like grids, and summarized on demand~\mbox{\cite{webstrates,bederson1994pad,wright2006sandbox,wildcard2020litt}}. By breaking the walls of individual webpages, this malleable workspace lets users keep all relevant pages for a task in view while flexibly reorganizing them as goals evolve. AI facilitation further enables ``orchestration,'' where users organize, manipulate, and synthesize content across webpages with the help of AI agents in a scalable, flexible manner.}

\rv{To ground our design exploration, we synthesize literature that identifies both the merits of manual effort and its cognitive costs in web browsing and sensemaking for information foraging and synthesis tasks~\mbox{\cite{pirolli1995information,pirolli2005sensemaking}}. Reflecting on the activities involved---from starting to synthesizing---we break down browsing-at-scale into specific interaction challenges and opportunities.}

\rv{We prototype the \mbox{\sys} browser to instantiates this malleable workspace. Instead of isolated tabs, \mbox{\sys} places pages side by side on a \emph{canvas}, where users can open large batches of links to broaden exploration, deploy autonomous agents across tabs to accelerate repetitive interactions, transform webpages into succinct representations, flexibly organize them into structures based on content and context, and synthesize information across pages on demand to support progressive sensemaking.
\mbox{\sys} leverages AI facilitation and interactive features to help users build and manage a dynamic information space, with the web as its foundation.}

To understand how effectively \sys supports user-driven browsing, we conducted a lab study with eight participants.
We found that \sys not only accelerated web exploration, but also encouraged broader and more confident foraging by reducing the overhead of managing context across pages.
Participants valued the parallel view for maintaining awareness and organizing tasks, and appreciated the ability to selectively delegate to AI while retaining control over information sources.
These findings suggest that \sys facilitates user-driven, at-scale browsing by aligning AI assistance with users' evolving goals, preferences, and sensemaking processes.

\section{Problem Framing}
\label{sec:background}

This work emerges amid rapid advancements in LLMs and their applications. Fully automated deep research agents like OpenAI Deep Research, Perplexity, and Manus now forage, synthesize, and generate comprehensive reports from simple user queries, transforming how we access information~\cite{venkit2024search,deepresearch,perplexity,manus}.

However, HCI and sensemaking research have long emphasized the benefits of the information foraging and synthesizing \emph{process}~\cite{rieh2016towards,teevan2004perfect}. Therefore, this work takes a different stance---We do not want to replace browsing with fully automated agents. Instead, we explore leveraging AI's capabilities and integrating them into each activity in the information-seeking and sensemaking process, to reduce cognitive and interaction costs while preserving the cognitive benefits of direct information engagement.

To capture users' informational needs and both the challenging and rewarding aspects of user-driven browsing, we review literature on information gathering, sensemaking, and web-based information tasks.
We structure our problem framing by adapting Ellis' behavioral model of information retrieval to include synthesis and transactional activities, seeking the full range of modern web tasks (Table~\ref{tab:bg})~\cite{ellis1989behavioural,ellis1993comparison,ellis1997modelling,choo1999information,chang2021tab,sellen2002knowledge}.
For each activity, we identify costs, highlight cognitive benefits of user-driven browsing, and propose opportunities for AI facilitation.

\begin{table*}[htp]
  \caption{Activities, Costs, and Benefits of User-Driven Browsing, and Opportunities for Facilitation}
  \label{tab:bg}
  \renewcommand{\arraystretch}{1.6} 
  \setlength{\tabcolsep}{4pt} 
  \small
  \centering
  \begin{tabular}{
    @{}>{\centering\arraybackslash}p{1.8cm}
    >{\centering\arraybackslash}p{2.5cm}
    >{\centering\arraybackslash}p{2.5cm}
    >{\centering\arraybackslash}p{3.1cm}
    >{\centering\arraybackslash}p{3.1cm}
    >{\centering\arraybackslash}p{3.1cm}@{}
  }
    \toprule
    \textbf{Activities} & \textbf{Tasks} & \textbf{Goals} & \textbf{Costs to Mitigate} & \textbf{Benefits to Preserve} & \textbf{Opportunities} \\
    \midrule\midrule
  \end{tabular}

  \begin{tabular}{
    @{}>{\raggedright\arraybackslash}p{1.8cm}
    |>{\raggedright\arraybackslash}p{2.5cm}
    |>{\raggedright\arraybackslash}p{2.5cm}
    |>{\raggedright\arraybackslash}p{3.1cm}
    |>{\raggedright\arraybackslash}p{3.1cm}
    |>{\raggedright\arraybackslash}p{3.1cm}@{}
  }
    \textbf{Starting} Section~\ref{sec:s} & Formulate query, select source & Establish an initial direction for search & Repeatedly translate goals to queries and compare results among different sources & Enable goal-oriented and anchored exploration & Bootstrap explorations through parallel starting points across sources \emph{at scale} \\
    \hline
    \textbf{Scouting} Section~\ref{sec:sf} & Navigate links and citations, scan search results and lists & Expand, reorient, and contextualize the scope of search & Handling multiple threads across isolated pages causes information overload and context loss & Enable contextual understanding and serendipitous discovery & Accelerate exploration and cross-reference through navigation and overview \emph{at scale} \\
    \hline
    \textbf{Filtering} Section~\ref{sec:sf} & Remove irrelevant pages or content, highlight and bookmark favorites & Identify and prioritize relevant items while discarding the rest & Navigate through a large amount of irrelevant information and materials & Refine understanding and criteria that lead to more accurate future queries & Support evaluating, comparing, and refining exploration results \emph{at scale} \\
    \hline
    \textbf{Revisiting} Section~\ref{sec:mc} & Organize and revisit content, track changes & Organize, track, and revisit useful content & Locating stacked pages and tracking scattered information are effortful and disorienting & Refresh memories of information and form new connections and ideas with updated context & Flexibly organize webpages \emph{at scale} to support refind and diverse task needs \\
    \hline
    \textbf{Collecting} Section~\ref{sec:mc} & Save, copy, or import information to workspaces & Capture, transform, and aggregate content across sources & 
    Saving information from multiple disconnected sources continuously disrupts workflow & Externalize understanding and build structured, curated knowledge & Extract and overview information across webpages \emph{at scale} \\
    \hline
    \textbf{Synthesizing} Section~\ref{sec:st} & Compare, integrate, and schematize information & Integrate and structure materials into a coherent understanding & Integrating fragmented content into coherent understanding is cognitively demanding & Connect insights across sources and deepen understanding & Dynamically synthesize web content across webpages \emph{as scale} \\
    \hline
    \textbf{Transacting} Section~\ref{sec:st} & Complete operations and communicating information & Act upon, interpret, and communicate the findings & Performing repetitive actions across interfaces is tedious and time-consuming & Build user agency and trust towards the final results & Streamline repetitive operations across pages \emph{at scale} through parallel, on-demand automation \\
    \bottomrule
  \end{tabular}
\end{table*}

\subsection{Starting}
\label{sec:s}

Informational tasks often start without well-defined goals and domain-specific knowledge and language that are needed to translate users' vague needs into concrete queries~\cite{aula2008complex,marchionini2006exploratory,hearst2009search,ellis1989behavioural}. As a result, selecting starting points and crafting initial queries can be demanding and often lead to confusion and ineffective paths, requiring users to iteratively adjust their queries and sources throughout the process~\cite{aula2008complex,bates1989design,white2009exploratory,kuhlthau1991inside}. On the other hand, being able to flexibly restart and reformulate queries allows users to adaptively anchor and reorient their exploration and gain a deeper understanding of their goal and necessary vocabulary~\cite{vakkari2001changes,rieh2014searching}.

Different approaches for query assistance and recommendation have been explored to help people formulate and reformulate queries to search more effectively, including assisting completion and refinement~\cite{palani2021conotate,lau1999patterns,agapie2013leading}, session-based search~\cite{radlinski2010inferring}, and related query recommendations~\cite{jones2006generating,kelly2009comparison,silvestri2009mining,williams1984makes}. Contextual information, such as browsing history and notes have been leveraged to generate relevant suggestions~\cite{palani2021conotate,chang2019searchlens}. However, prior work focuses mainly on query assistance within scoped search engines or sources. Despite reviewing results helping users clarify their goals and improve queries, they still have to choose from recommendations and perform each search manually and repeatedly.

In this work, we explore moving beyond simple query suggestions by enabling AI to proactively search multiple sources in parallel, helping users quickly access, overview, and compare results across sources to support broader exploration.

\subsection{Scouting and Filtering}
\label{sec:sf}

When an information task begins, users navigate through links, citations, and search results to expand or reorient their exploration~\cite{bates1989design,pirolli1999information,pirolli2005sensemaking,o1993orienteering,huberman1998strong}.
This exploration process helps build contextual understanding, broader topic awareness, and mental maps of the information space~\cite{suh2023sensecape,bates1989design,pirolli1999information}, enabling the users to begin to recognize relationships between concepts, sources, and navigation paths~\cite{aula2008complex,teevan2004perfect}.
It also creates opportunities for serendipitous discovery, where users encounter unexpected connections and insights~\cite{foster2003serendipity,agarwal2015towards}.

Scouting is tightly coupled with filtering~\cite{palani2022interweave,pirolli1999information}.
Users continuously assess relevance, track what they have seen, and decide what to keep or discard~\cite{marchionini2006exploratory,bates1989design,chang2021tab}.
These processes are demanding. Users must scan and manage parallel sessions while distributing attention and memory across scattered information across webpages~\cite{marchionini2006exploratory,bates1989design,pirolli1995information,von2013dobbs}.
Tabbed browsers amplify the burden. They present webpages as a flat, linear stack, requiring users to view them in isolated viewports and switch back and forth to compare and cross-reference~\cite{chang2021tab,chang2021tabs,dubroy2010study,brown1995deckscape}. 
Following each page manually can also be costly and discourage exploration~\cite{huberman1998strong,min2025malleable}.
As exploration progresses, maintaining context or spotting connections becomes increasingly difficult~\cite{ma2023browsing,chang2021tabs,chang2021tab,weinreich2006off}.

Prior work explored visualizing information scent, traces, and relationships among webpages~\cite{cockburn2000issues,olston2003scenttrails,ayers1995using,lamping1995focus,trigg1988guided,dork2008visgets}, helping manage browsing progress~\cite{newfield1998scratchpad,chang2021tabs}, and providing assistive browsing support, such as recommendations, to promote exploration~\cite{lieberman1995letizia,rhodes2000just,rhodes1996remembrance}. However, users still have to drill down into each page manually, one by one, to review content and complete navigation, making the process slow and repetitive.

We argue that information and interactions on the web are fundamentally distributed, and the browser must evolve beyond supporting isolated, single-page workflows~\cite{ma2023browsing,brown1995deckscape}.
Instead of leaving users to manually manage and constantly switch among different pages, we explore supporting viewing web content from multiple pages at the same time and operating on them in parallel, to help comparison, maintain context across pages, and follow multiple exploration threads simultaneously.

\subsection{Revisiting and Collecting}
\label{sec:mc}

As exploration grows more complex, the need to revisit information increases, but doing so becomes more demanding and time-consuming.
Revisiting information helps refresh memory, preserve context, track updates, and form new connections and ideas~\cite{adar2008large,tauscher1997people,kittur2013costs,capra2003re,morris2008searchbar,jones2001keeping}. However, this process can be costly in web browsing. Users may keep related tabs or bookmarks, which quickly accumulate and lead to disorganized spaces and an increasingly difficult refinding process~\cite{chang2021tab,chang2021tabs,ma2023browsing}. If a webpage is closed, current browsers poorly support refinding visited content, as they only record links and page titles~\cite{chang2021tab}.

An important approach to facilitate revisiting is to collect information into a dedicated personal workspace, like a document or folder, and reorganize it according to needs and interpretations~\cite{kittur2013costs,pirolli1995information,pirolli2005sensemaking}.
This process externalizes intermediate thinking and builds structured understandings of the information space~\cite{kittur2013costs,pirolli1995information,pirolli2005sensemaking}. However, it requires significant manual effort to transform information and restructure the collections, disrupting workflow~\cite{liu2022wigglite,kittur2013costs}.

Prior work has explored supporting the collection of information on the web by enabling the capture of finer-grained units~\cite{schraefel2002hunter,kittur2013costs}, reducing collection efforts~\cite{liu2022wigglite,stylos2004citrine,chang2016supporting}, automating collection or organization~\cite{leetiernan2003two,won2009contextual,liu2022crystalline,chang2021tabs}, contextualizing exploration with collected information~\cite{palani2022interweave,rachatasumrit2021forsense}, and enabling in-situ and progressive sensemaking through growing collections~\cite{liu2019unakite,kuznetsov2022fuse}.

In this work, we shift to treating the \emph{browser} as the primary information workspace, where users can revisit, collect, and organize multiple webpages simultaneously. Webpages can be flexibly structured, extended, and abstracted into simpler formats for faster scanning and easier refining.

\subsection{Synthesizing and Transacting}
\label{sec:st}

After gathering information, users compare and integrate materials to form a coherent understanding. They organize evidence, generate insights, and identify gaps in their knowledge~\cite{pirolli2005sensemaking}. This process helps users refine their mental models and create structured interpretations of the information~\cite{pirolli2005sensemaking,russell1993cost,pirolli1999information,liu2019unakite}.

Synthesizing is cognitively intensive. Users must track multiple representations, manage conflicting or redundant information, and translate unstructured content into coherent frameworks~\cite{pirolli2005sensemaking,kittur2013costs}.
Synthesizing is often gradual and interleaved with foraging~\cite{pirolli1999information}.
However, in traditional web browsing, this task often occurs in a disconnected space from browsing and exploration: users switch to separate tools---documents, spreadsheets, or note-taking apps---to build structured outputs. This fragmentation loses context and interrupts workflows~\cite{kittur2013costs,liu2022wigglite}.

Task-oriented decision-making and operations are increasingly prevalent in web activities~\cite{chang2021tabs}. Meanwhile, navigating and operating the highly interactive web are essential to many of the previously mentioned activities, from scouting to filtering.
However, users often face tedious and repetitive low-level operations that slow down the exploration and hinder flexible synthesis~\cite{bolin2005automation,pu2023dilogics}.

Meanwhile, web and UI automation agents powered by LLMs show increasing promise in reducing manual efforts in web browsing~\cite{zhou2023webarena,browser2024,he-etal-2024-webvoyager,koh2024visualwebarena,deng2023mind2web,huq2025cowpilot}. These agents, including OpenAI Operator and Claude Computer Use, translate natural language instructions into concrete actions (like clicks and scrolling), continuously analyze unseen interfaces, and adaptively determine the next steps to take without requiring extensive end-user training, scripting, or step-by-step monitoring~\cite{openaioperator,computeruse}. While their performance in the wild still needs substantial improvement, the rapid advancement of LLMs suggests we are stepping into a promising future where automation agents may reliably and swiftly navigate convoluted web environments, gather information, and complete complex tasks.

In this work, we explore on-demand automation and flexible, progressive synthesis. By displaying multiple webpages side by side, users can monitor multiple web automation agents and control repetitive operations across pages in parallel. We also leverage the synthesis capabilities of LLMs to dynamically generate digests that adapt to changing selections and evolving page content, helping users understand information at scale.

\subsection{Related Work}

We briefly review other system work that our research builds on.

\subsubsection{Canvas- and spatial-based interfaces}

\rv{Our work builds on a long line of systems that use spatial canvases and multi-workspace environments to help people manage complex information. Early systems such as Rooms, Pad++, and BumpTop explored multiple virtual workspaces, zoomable canvases, and physics-inspired desktops to cope with limited screen space and window thrashing~\mbox{\cite{henderson1986rooms,bederson1994pad,agarawala2006keepin}}. MosaicG visualized web history as a two-dimensional map of visited pages to support revisitation and orientation~\mbox{\cite{ayers1995using}}. Code Bubbles and other working-set interfaces showed how collections of lightweight fragments arranged on a canvas can improve code understanding and reduce navigation overhead~\mbox{\cite{bragdon2010code}}. Tools such as Figma demonstrate how shared canvases for many artifacts support parallel, non-linear work in everyday practice~\mbox{\cite{figma}}. Recent LLM-based systems such as Graphologue and Sensecape further leverage spatial layouts and multilevel structures to externalize, reorganize, and explore model-generated content~\mbox{\cite{jiang2023graphologue,suh2023sensecape}}.}

\rv{These efforts demonstrate the value of spatial canvases for organizing many items and supporting non-linear exploration. In this work, we explore embedding webpages on a canvas and coupling spatial arrangement with AI-facilitated operations across them, enabling parallel automation and synthesis at the browser level.}

\subsubsection{Malleable software and web interfaces}

\rv{Malleable software research has long envisioned systems that users can adapt to their own tasks, data, and workflows~\mbox{\cite{pdm}}. Haystack proposed unified data models and customizable interface layers to integrate heterogeneous information and support personalized information management~\mbox{\cite{adar1999haystack}}. Smalltalk enabled users to inspect and modify the behavior and appearance of GUI objects, and Webstrates treated web documents as shared, dynamic media that can be restructured and recomposed across clients~\mbox{\cite{webstrates,kay1996early}}. More recently, systems such as Jelly explore activity-based, data-model-driven customization, where users can reshape information interfaces through natural language and direct manipulation as their tasks evolve~\mbox{\cite{cao2025jelly}}. Work on malleable web interfaces further argues for treating websites as adaptable surfaces that can be reshaped to better fit user goals and cross-site workflows~\mbox{\cite{min2025malleable,wildcard2020litt}}.}

\rv{In this work, we treat webpages as malleable materials in a dynamic workspace for the users to forage and make sense of web information at scale. AI agents extract, synthesize, and operate across many sites in parallel to enable users' interaction with underlying webpages and orchestration through the malleable workspace.}

\section{Summary and Design Goal}

\rv{Web information tasks involve fluidly interleaved activities from starting and scouting, to synthesizing and transacting across webpages. We synthesize key insights from prior literature in Table~\mbox{\ref{tab:bg}}, outlining typical tasks, user goals, associated costs and benefits, and opportunities for facilitation in this information foraging and sensemaking process.}

\rv{Prior research has highlighted the value of manual effort in these activities: it supports anchored exploration aligned with evolving and personal needs~\mbox{\cite{white2009exploratory,ellis1989behavioural,teevan2004perfect}}; enables serendipitous discovery~\mbox{\cite{foster2003serendipity}}; and fosters progressive understanding of the problem space~\mbox{\cite{teevan2004perfect,suh2023sensecape,pirolli2005sensemaking}}. Manual interaction also contributes to building reusable knowledge, understanding the reasoning behind final results, and maintaining user agency and trust~\mbox{\cite{teevan2004perfect,pirolli1995information}}.}

\rv{Meanwhile, web-based tasks require users to repeatedly navigate through distributed webpages and manage large volumes of fragmented information and context. Traditional tabbed browsing, however, prioritizes working within each individual page and barely supports cross-page activities, amplifying navigation and synthesis costs~\mbox{\cite{chang2021tab,ma2023browsing,brown1995deckscape}}.}

\rv{Therefore, this work aims to facilitate a user-driven approach to the web information space that preserves the merits of manual effort while reducing the cognitive and interaction costs imposed by the distributed nature of web-based tasks. We target information-foraging and sensemaking workflows that span tens of webpages, and seek to support them through a malleable browser-level workspace for \mbox{\textbf{browsing at scale}}.}

\begin{figure*}[ht]
    \centering
    \includegraphics[width=\textwidth]{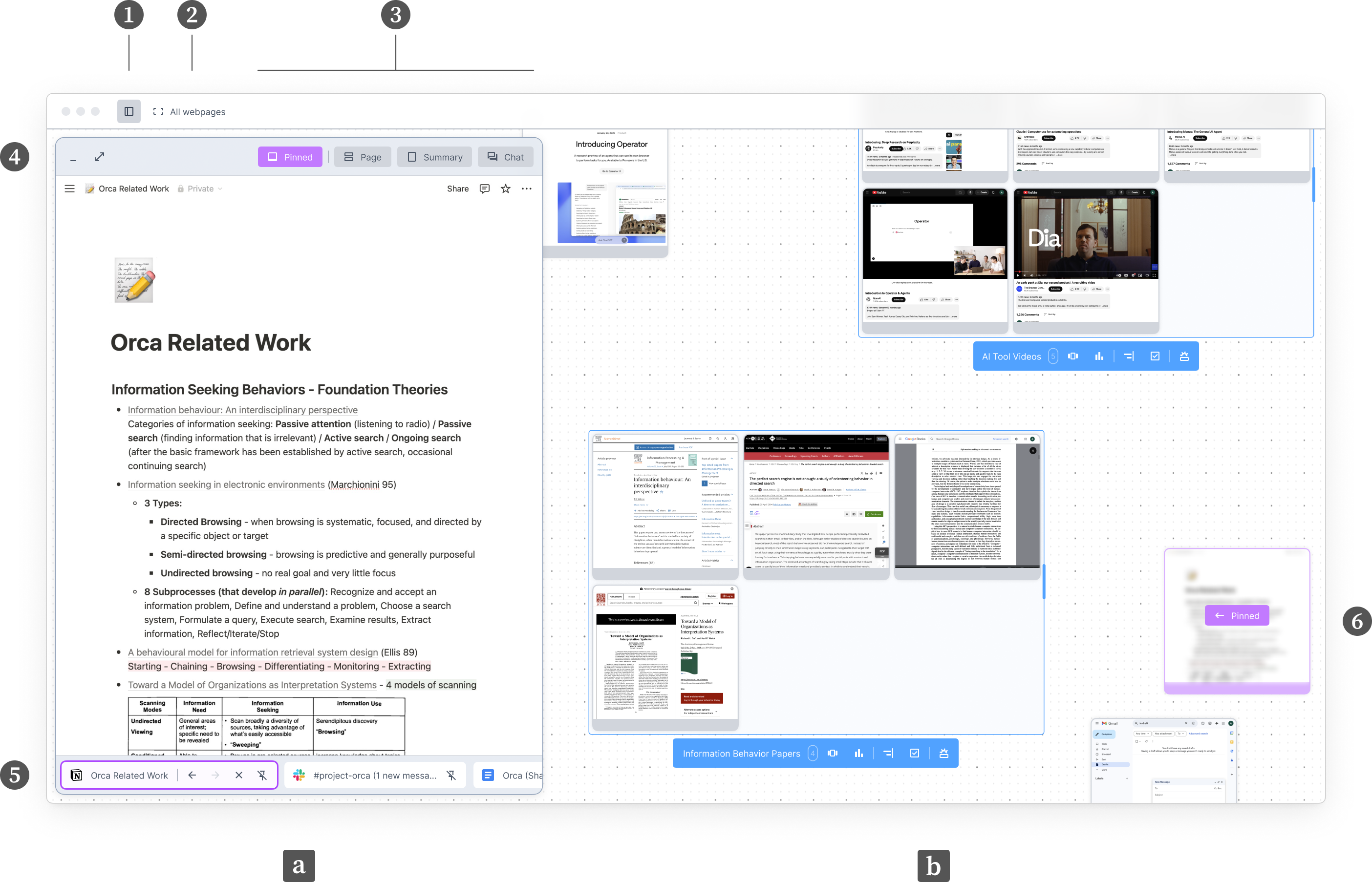}
    \caption{The \sys interface features a \webcanvas (b), where webpages can be spatially arranged and organized, and a side panel (a), which can be toggled (1) or set to fullscreen (4). The side panel shows pinned pages or summaries---either for individual webpages or multiple selected ones (3). Users can zoom out the viewport to view all pages at once (2). When a webpage is pinned, its instance on the canvas turns purple (6). Users can pin a set of pages and switch among them (5).}
    \Description{A screenshot of the \sys browser's two-part user interface. The interface is divided into a narrow, foldable side panel on the left and a large Web Canvas on the right. The side panel shows a website and a row of tabs at its bottom; one tab is highlighted in purple to indicate it is currently selected. The Web Canvas is a freeform canvas where multiple webpage windows are spatially arranged and grouped.}
    \label{fig:orca}
\end{figure*}

\subsection{Design Goal: Browsing at Scale}

\rv{We aim to enable browsing at scale to reduce repetitive low-level operations and constant context switching in cross-page tasks. In our work, browsing at scale means that users can treat a large set of related pages as a single, malleable workspace and:}

\begin{itemize}[label=$\circ$, leftmargin=2em]
  \item Instead of repeatedly restarting search with new queries or drilling into one thread or subpage at a time, users can \textbf{start and navigate at scale} by expanding from their current context to surface many relevant resources and explore multiple pages in parallel.

  \item Instead of clustering pages in linear, cluttered stacks and jumping back and forth between tabs, users could \textbf{view and organize at scale} across multiple pages at the same time.

  \item Instead of finding information from each page individually, users could \textbf{extract at scale} by surfacing key content across many pages simultaneously.
  
  \item Instead of repeating low-level actions inside individual pages, users could \textbf{operate at scale} by broadcasting intents and operations across multiple pages in one action.

  \item Instead of making sense of information from fragmented places manually, users could \textbf{synthesize at scale} with summaries that integrate content of multiple sources.
\end{itemize}

\section{Approach}

\rv{We explore a design approach that supports \mbox{\textbf{browsing at scale}} by (1) reconceptualizing webpages as ``malleable materials'' and the browser as a ``malleable space'' that together form a task-centric workspace~\mbox{\cite{webstrates}}, and (2) employing AI as a facilitator that scales users' reach, extraction, and operation across that workspace while keeping them in control as an ``orchestrator.''}

\rv{First, we address the limitations of traditional tabbed browsing, which treats each page as an isolated primary workspace organized in a linear stack, making it difficult to view or operate across multiple pages at once and forcing users to repeatedly sift through unrelated content. We explore treating webpages as malleable materials that can be viewed at varying scales in parallel, selectively extracted into compact representations, and reorganized into structures such as grids, stacks, and tables (Sections~\mbox{\ref{sec:overview}}--\mbox{\ref{sec:structures}}). This malleable browser space lets users arrange, group, and recombine pages into custom workspaces aligned with their tasks, so that cross-page comparison, filtering, and sensemaking happen in one shared context rather than across scattered tabs.}

\rv{Second, we support user-driven, AI-facilitated orchestration across this malleable workspace. Instead of end-to-end automation that removes users from the loop, AI helps \mbox{\emph{scale up}} exploration by expanding from the current context to related sources and alternatives (Section~\mbox{\ref{sec:dm}}), \mbox{\emph{surface}} structure by extracting fields and synthesizing summaries across pages to aid sensemaking (Sections~\mbox{\ref{sec:extraction}}--\mbox{\ref{sec:summarization}}), and \mbox{\emph{execute}} repetitive low-level operations in parallel across selected pages (Section~\mbox{\ref{sec:agent}}). Users remain responsible for \mbox{\emph{scaling down}}---filtering, organizing, deep-diving, and steering automation---so that browsing at scale emerges as a collaborative process between user and AI over malleable webpages, rather than as fully delegated web automation.}

\section{Browsing at Scale with \sys}
\label{sec:orca}

\rv{We present \mbox{\textbf{\sys}}, a browser designed to support user-driven web information foraging at scale (Figure~\mbox{\ref{fig:orca}}).}

\begin{scenario}
\rv{To illustrate how \mbox{\sys} streamlines complex browsing workflows, we follow \textbf{Leela}, a user planning a family trip to Paris. Through her journey, we describe how \mbox{\sys} transforms the isolated webpages into a malleable workspace for gathering, organizing, and synthesizing information.}
\end{scenario}

\subsection{View at Scale: Spatial Overview Across Multiple Webpages}
\label{sec:overview}

\begin{figure}
    \centering
    \includegraphics[width=\linewidth]{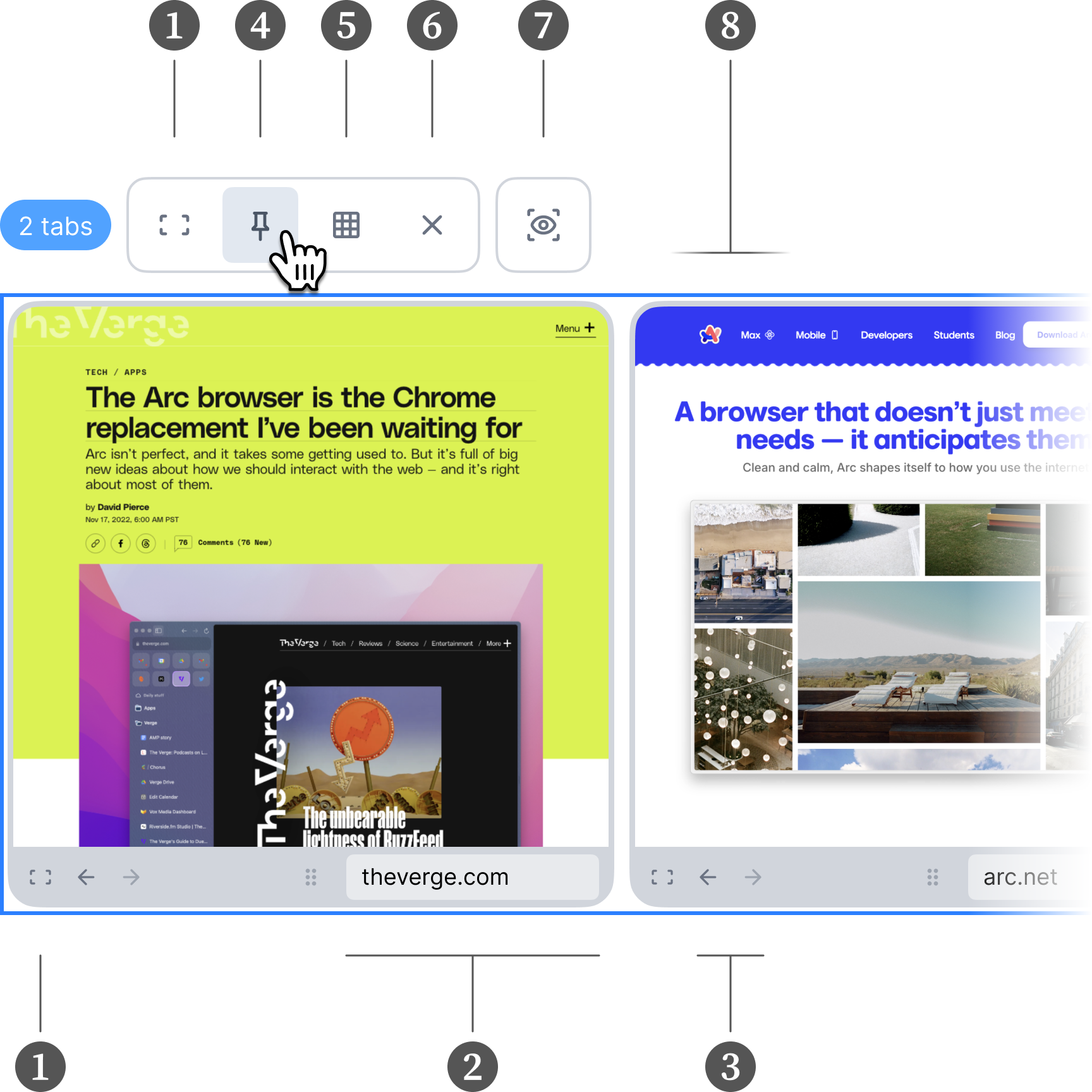}
    \caption{In \sys, webpages can be arranged side by side. Users control individual pages via bottom controls---e.g., update URL (2) or navigate back or forward (3)---like in traditional browsers. They can also act on selected pages using the contextual menu (top), indicated by the selection box (8). The viewport can be adjusted to show one or multiple selected pages (1). Pages can be pinned to the sidebar for static view (4), arranged in a grid (5), batch closed (6), or extracted via the \extraction toggle to surface key information (7).}
    \Description{Two webpages are shown side-by-side in the \sys interface. Each webpage has its own set of controls beneath it, including back and forward navigation buttons and a URL bar. Above the pages is a shared contextual menu with icons for actions that apply to both selected pages, such as pinning them or arranging them in a grid. A cursor is shown pressing the `Pin to the sidebar' button.}
    \label{fig:webpage}
\end{figure}

\begin{scenario}
\rv{Leela plans a family trip to Paris. She searches for hotels and drags candidates onto the \mbox{\webcanvas}. With \sys, she places family-friendly options at the top, groups budget ones on the left, and moves backups into a corner. She zooms out to see all options, then zooms in to read a few reviews in detail. She later sees that some hotels do not meet her criteria---she range-selects them and closes them all at once.}
\end{scenario}

\rv{Traditional tabbed browsing fragments context and content across a linear sequence of pages, making it difficult to maintain context when web information branches and interweaves.
To address this, \mbox{\sys} supports showing multiple webpages in parallel by displaying them as ``shapes'' on the \emph{\mbox{\webcanvas}} (Figure~\mbox{\ref{fig:orca}}.b).}

\rv{Being able to easily build and maintain a view of many pages side by side helps users understand, track, and synthesize information at scale. As a result, users zoom, pan, and lay out webpages in ways that reflect their ongoing tasks within an ad hoc, customized workspace that better fits their ongoing task, for example, grouping hotels by price, flights by stops, or papers by topic. While some content may appear smaller and harder to read or operate on, \mbox{\sys} supports extracting, operating on, and synthesizing this content with AI facilitation, which we introduce in the following subsections. This allows users to utilize \mbox{\webcanvas} to forage and orchestrate at a higher level.}

\rv{On the other hand, because \mbox{\sys} shows the actual content of each page rather than only titles or thumbnails, users rely on rich visual cues for refinding and comparison. Range selection and batch actions enable users to reposition or act on many pages at once~\mbox{\cite{masson2024textoshop,pdm}}. Dragging a link from a page onto the canvas opens it at a chosen position, while clicking opens it adjacent to the source page for immediate side-by-side inspection~\mbox{\cite{vistrates,beaudouin2012multisurface}}.}

\subsubsection{Pinning webpages for fixed and concentrated views}

\noindent
\begin{scenario}
\rv{As Leela narrows down hotels, she starts to note them down in Google Docs. She pins this note page to the left sidebar so it stays full size, then continues to rearrange, open, and close hotel pages on the \mbox{\webcanvas}. Whenever she finds a promising option, she glances at the pinned note, updates her shortlist, and hides the sidebar when she wants a distraction-free canvas view.}
\end{scenario}

\rv{For any webpage that users want to keep in a fixed, prominent view (for example, a document they are editing or a reference page), they can \emph{pin} it to the side (Figure~\mbox{\ref{fig:webpage}}.4). Pinned webpages are displayed at their original size, as in a traditional browser, and can be switched via the \emph{\mbox{\pinnedbar}} (Figure~\mbox{\ref{fig:orca}}.a). Unlike a traditional tab bar---which shows every open tab and forces users to manage them all at once---\mbox{\sys} lets users create dynamic tab bars that hold only selected webpages, while other pages remain open on the \mbox{\webcanvas}.}

The pinned sidebar can be used together with the underlying canvas, allowing the user to leverage both the static view of a ``base'' page and the dynamic, exploratory nature of the \webcanvas.
For example, they may pin a webpage of a notetaking app on the side while continuing to explore information using the \webcanvas on the right. The pinned sidebar can also be hidden or expanded fullscreen for a concentrated view of only the pinned pages.

\subsection{Organize at Scale: Flexibly Organizing Webpages with Different Structures}
\label{sec:structures}

\begin{scenario}
\rv{After twenty minutes of exploring, Leela has dozens of hotel pages scattered across the \mbox{\webcanvas}. She range-selects them and asks \mbox{\sys} to organize them into a grid for a clean overview. After sorting the options based on her preferences, she switches the grid into a stack to set aside before exploring flights.}
\end{scenario}

\rv{Manually arranging webpages on a freeform canvas and managing the viewport can be demanding. Unlike positioning shapes in Figma, where layout is part of the design task, such effort can feel like an added burden~\mbox{\cite{kosch2023survey}}. A cluttered canvas also makes it harder to discover connections, synthesize information, or cross-reference pages. To address this, \mbox{\sys} provides ways to gather, tidy, and recompose webpages into different structures.}

\begin{figure}
    \centering
    \includegraphics[width=\linewidth]{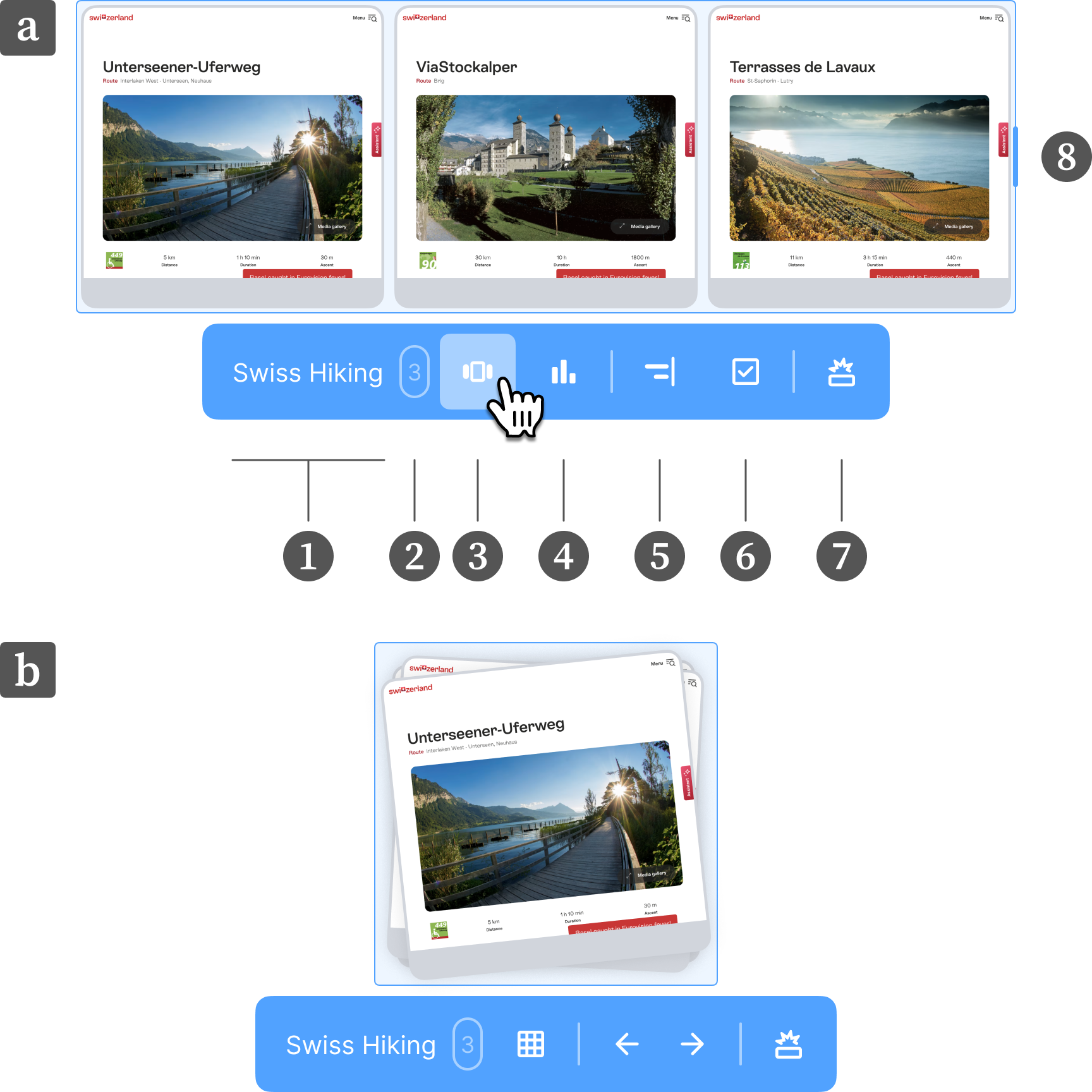}
    \caption{Webpages can be organized into a grid (a) or stack (b) layout. Upon creation, \sys assigns the group a short, customizable name based on page content (1), along with its total page count (2). Users can switch between grid and stack layout (3) or convert the group into a table or visualization (4). Pages can be sorted (5) or selectively chosen (6) based on custom criteria. Users can adjust the number of columns in the grid layout to control its width (8). When no longer needed, groups can be dissolved (7).}
    \Description{This figure contrasts two organizational layouts for a group of webpages. The top image shows a grid layout, with three webpages displayed side-by-side. The bottom image shows a stack layout, where only the first page of the group is visible. Both views feature a shared control bar below the pages that displays the group's name and page count, along with icons to manage the group. A cursor is shown clicking the button to switch from the grid to the stack layout.}
    \label{fig:organizations}
\end{figure}

\rv{\mbox{\sys} offers \emph{grid} and \emph{stack} organizations. Users can place pages in a grid for a clean, scannable layout (Figure~\mbox{\ref{fig:organizations}}.a) and sort the grid by content or browsing metadata (such as time opened or last interacted) to reveal relationships or locate pages quickly. To archive content while keeping tabs available, or for pages not in active use, users can place them in stacks and flip through them as needed (Figure~\mbox{\ref{fig:organizations}}.b). Webpages can be dragged in or out of grids and stacks.}

Prior work has explored representing multiple webpages as diagrams or nested lists to show their relationships~\cite{ayers1995using,chang2021tabs}.
\sys explores representing webpages with tables and interactive graphics~\cite{logit}.
\sys displays content from each page in columns of a table.
Users can continue adding more columns to extract more information across all pages in the table.
Webpages can also be visualized as interactive diagrams (e.g., bar charts) using Vega-Lite grammar generated on demand for visualizing specific aspects of content across pages, e.g., when comparing hotel prices~\cite{2017-vega-lite}.

\subsection{Navigate at Scale: Batch Exploration in Depth and Breadth Through Direct Manipulation}
\label{sec:dm}

\begin{scenario}
\rv{With the hotel stack set aside, Leela opens a flight search page for her trip dates. Instead of going through the results one by one, she uses \mbox{\batchopen} capsule to filter and open all nonstop flights that match her preferred departure window into a grid. From one flight detail page, she then pulls out the \mbox{\expansion} handle to explore a few alternative routes from other airlines into the same region of the canvas, dragging the handle to reserve space for just a handful of extra options.}
\end{scenario}

\begin{figure}
    \centering
    \includegraphics[width=\linewidth]{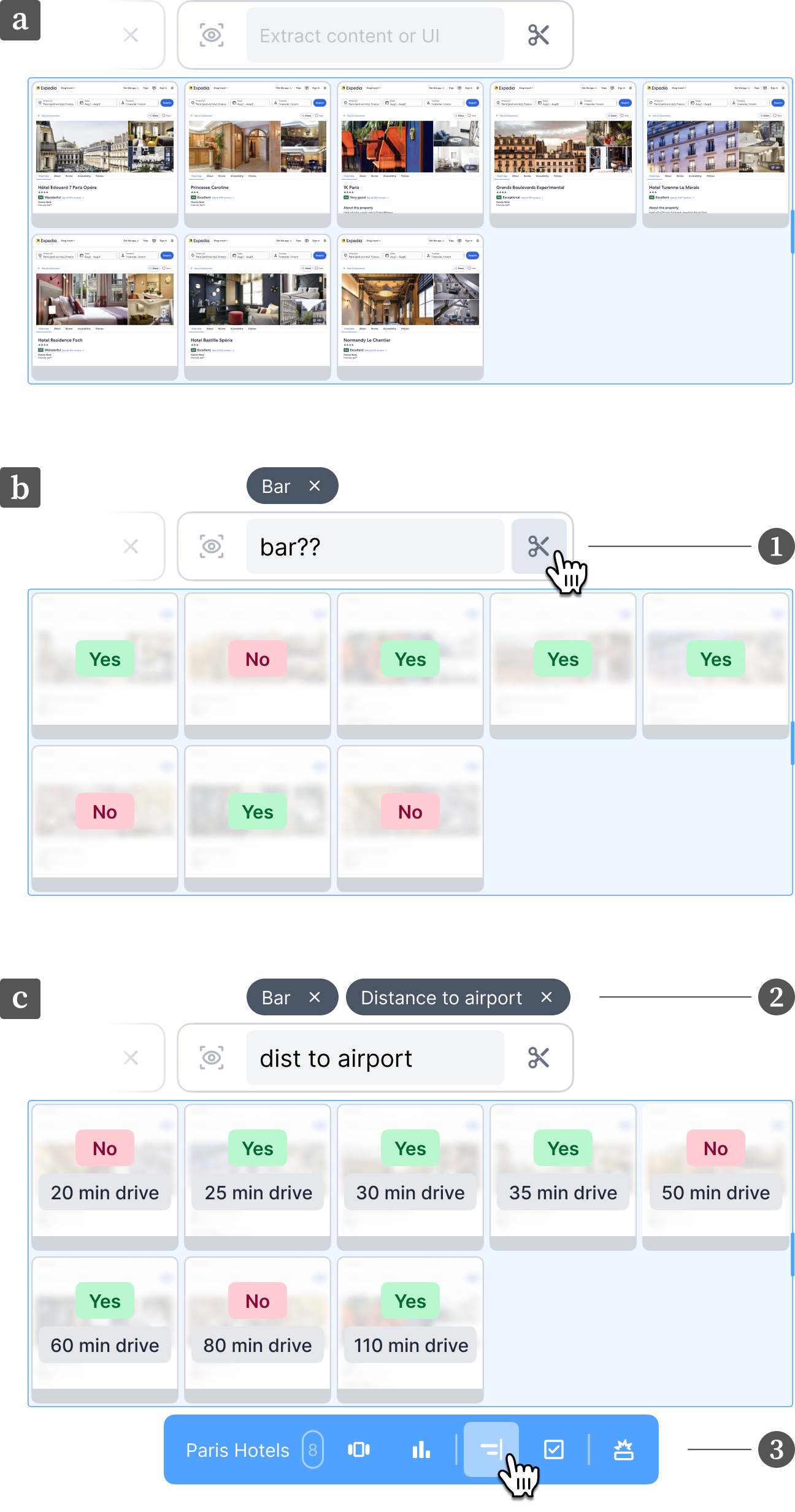}
    \caption{Users can extract and surface information across pages at scale for quick scanning. Users start by selecting a set of webpages (a), entering a simple extraction query (b), and pressing the \emph{Extract} button (1). \sys then displays the extracted information in a consistent format across pages. Users can continue customizing the views of pages to surface the information they care about (c). Extraction queries are saved (2), and hovering over a query highlights the corresponding extracted information for all pages. Users can further close or sort pages (3) to continue exploration.}
    \Description{A three-part diagram shows the process of extracting information across multiple webpages. The process begins with a grid of hotel webpages. Next, after the user queries for `Bar', the view changes so each webpage preview is covered by a simple card showing `Yes' or `No'. In the final step, another query is input by the user for the `Distance to airport', and the cards in the grid update to show both pieces of information for each hotel. A cursor is shown clicking the button to change the webpage grid to the chart view.}
    \label{fig:extraction}
\end{figure}

\begin{figure*}
    \centering
    \includegraphics[width=\textwidth]{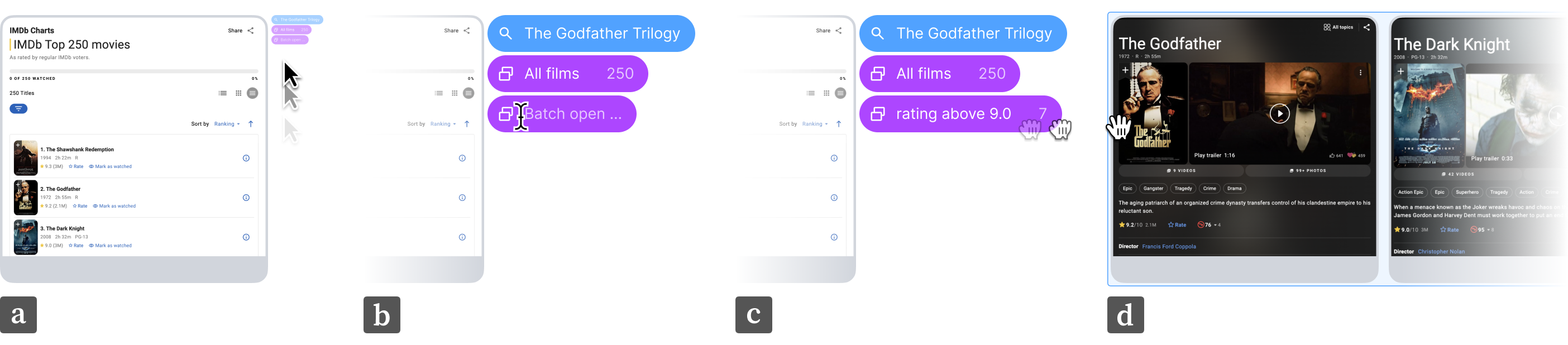}
    \caption{\sys facilitates in-depth exploration by allowing users to batch open multiple links from a webpage. The \batchopen menu appears minimized by default at the top right of selected pages (a). On hover, users can pick from suggested queries or enter a custom one (b). \sys then searches the page and compiles matching links, showing the total count to support query refinement (c). Users can drag the capsule onto the \webcanvas to open all links, automatically arranged in a grid (d).}
    \Description{A four-part diagram demonstrates a feature for opening multiple links from a webpage at once. The process starts with a user hovering over the `sidecar' menu on a movie-ranking webpage, minimized to avoid distraction, which enlarges and reveals a list of suggested queries to extract and open links in the page. The user then inputs a custom query of `rating above 9.0', dragging the purple capsule that represents the 7 matching links onto the Web Canvas, which automatically opens all 7 movie pages and arranges them in a grid.}
    \label{fig:batchopen}
\end{figure*}

Users often explore both in depth and breadth during an information task, in a highly interleaved manner.
However, they currently have to manually click each link on a page to drill down, go back to the base page to repeat the process (e.g., for a search results page); or manually switch between threads and re-enter queries to expand their exploration (e.g., to find results for alternative search queries).

The tedious process and the effort required to manage context across multiple threads often discourage users from exploring or fully utilizing different threads.
Therefore, \sys streamlines exploration in both directions of depth and breadth through direct manipulation within \webcanvas.

\rv{To help users quickly drill down and preview many links, \mbox{\sys} can \emph{\mbox{\batchopen}} links, such as \emph{``Discounted hotels''} on a listing page (Figure~\mbox{\ref{fig:batchopen}}). Users open these links by dragging the capsule button onto the canvas. They can also define custom extraction queries, such as \emph{``Hyatt, rating above 9''}. \mbox{\sys} extracts matching detail pages, shows the total number found, and lets users drag the capsule or press \mbox{\texttt{Return}} to open them on the side.}

\begin{figure*}
    \centering
    \includegraphics[width=\textwidth]{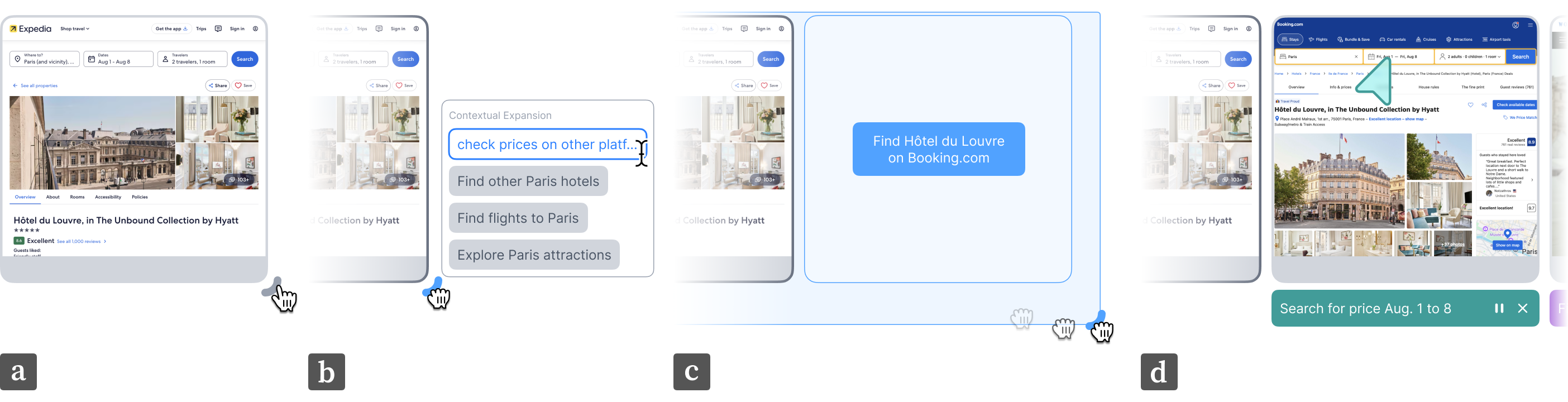}
    \caption{\sys supports in-breadth exploration through \expansion, which opens relevant pages based on context and user input. When a webpage is selected, a small handle appears at its bottom right corner (a). Clicking it reveals the \expansion menu, where users can pick suggested queries or enter a custom one to scout related pages (b). Users can drag the handle to spatially set how many pages to add (c), or double-click to let the AI decide. \sys then opens the relevant pages; if a page is not directly accessible via a link, a web automation agent completes the navigation (d).}
    \Description{A four-part illustration demonstrates the `Contextual Expansion' feature. The process begins with a user clicking a small handle on a hotel webpage, which opens a menu of suggested queries, like checking prices on other sites. The user then drags the handle to designate space for the new page. In the final step, a new webpage from a competing travel site opens next to the original one. A status bar indicates that an automation agent is navigating the new page to find the requested information.}
    \label{fig:expansion}
\end{figure*}

To expand exploration in breadth, \mbox{\sys} suggests relevant pages based on context. After selecting one or more pages, users click the small handle in the bottom-right corner to open the \expansion menu (Figure~\mbox{\ref{fig:expansion}}). \mbox{\sys} analyzes the selection and proposes queries. For example, when the user selects a hotel page, it may suggest \emph{``Find flights''} or \emph{``Find on other websites''}. Users can also enter custom queries. They can then drag the handle to control how many results to open or press \texttt{Return} to let AI choose. When a target page cannot be reached directly, \mbox{\sys} deploys a web automation agent to navigate from a starting point (e.g., Google) to the desired page (Section~\mbox{\ref{sec:agent}}).

\subsection{Extract at Scale: View Customization Through Continuous Page Extraction}
\label{sec:extraction}

\begin{scenario}
\rv{After shortlisting flights, Leela brings her hotel stack back into a grid to check their distances from the airport where her flight lands. She selects the hotel grid and turns on \mbox{\extraction}, adding the prompt ``Distance to ORY.'' \mbox{\sys} overlays each hotel page with the distance provided by the detail page. She quickly removes hotels that are too far from the airport and keeps a small set of viable options.}
\end{scenario}

\rv{Our prototype shows that users can still read and interact with web content at about one-third scale on a 14-inch monitor, fitting up to nine pages at once. However, it is critical to surface key content from each page at a familiar, constant size, regardless of the zoom level on the \mbox{\webcanvas}. Many pages also contain clutter that does not relate to the user's task. Helping users spot relevant content across pages can greatly improve exploration.}

\rv{To support focused access, \mbox{\sys} lets users extract content from webpages (Figure~\mbox{\ref{fig:extraction}}). Users select pages, toggle \mbox{\extraction}, and enter a query to pull out relevant information. For example, when comparing hotels, they might ask \emph{``Does it have a pool?''} or simply \emph{``Pool''}. \mbox{\sys} uses the LLM to interpret the intent, find answers on page, and display them in a clear, scannable format. The system also highlights the source on each webpage so users can inspect it when needed. This process allows more flexible filtering than the built-in tools on most sites~\mbox{\cite{min2025malleable}}.}

\rv{Users can also surface interactive elements to perform simple actions across pages directly from the \mbox{\webcanvas} overview. Instead of extracting raw code, \mbox{\sys} provides standardized UI elements---such as buttons, input boxes, and text areas---that represent the underlying widgets, allowing consistent UI extraction and generation~\mbox{\cite{cao2025jelly}}. The extracted components stay bi-directionally synced with the original pages through two mechanisms: content listeners update the extracted UI when the page changes, and web automation agents send user actions back to the page (Section~\mbox{\ref{sec:agent}}).}

\subsection{Synthesize at Scale: Dynamic Summary and Information Query}
\label{sec:summarization}

\begin{scenario}
\rv{Once she has a small set of flights and hotels on the \mbox{\webcanvas}, Leela selects them and opens the \mbox{\expose} sidebar. \mbox{\sys} shows a brief summary for each option and highlights key trade-offs, such as which combinations fit her budget and commute constraints. She then types a follow-up question asking which pairings best match her schedule. She skims the updated summary while rearranging pages on the canvas.}
\end{scenario}

\rv{Maximizing pages one by one and reading each in full can slow exploration. It also discourages users from opening and learning from the many webpages that \mbox{\sys} helps them create and manage.}

\begin{figure}[ht]
    \centering
    \includegraphics[width=\linewidth]{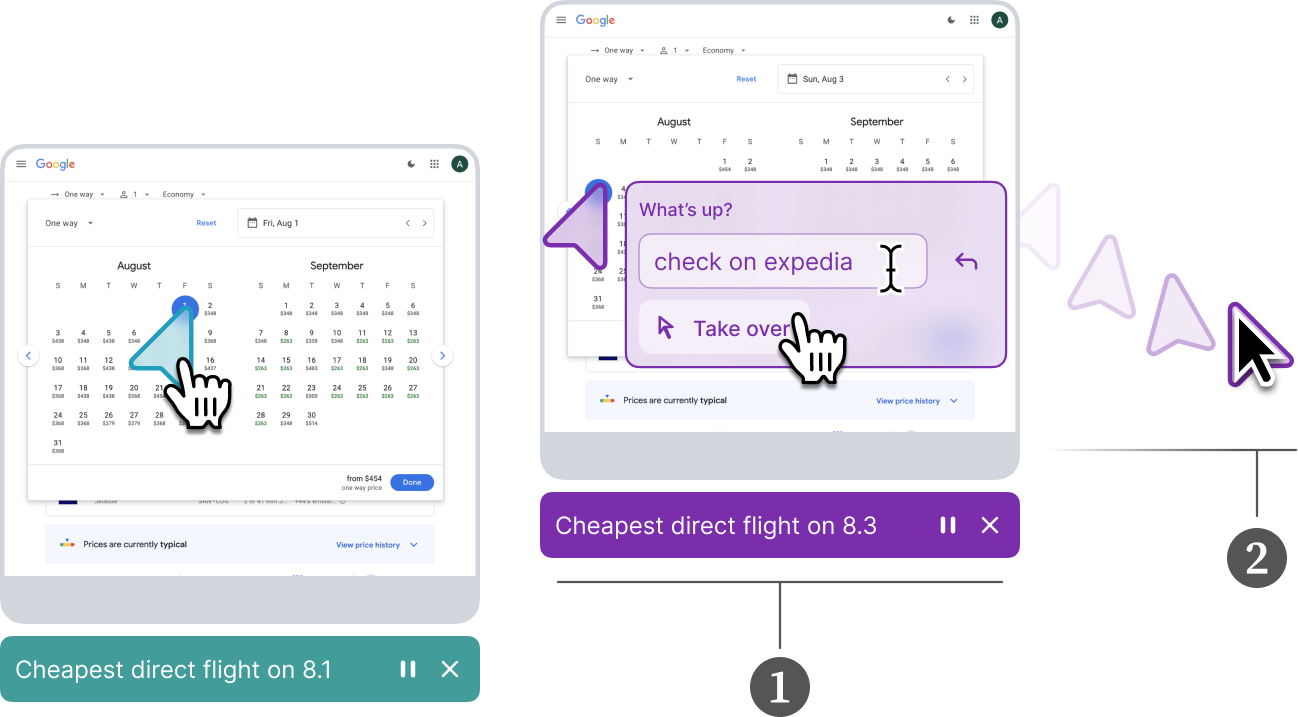}
    \caption{Multiple automation agents can operate in parallel on their respective webpages, with color-coding for visual tracking and distinction. Each agent has a control bar under the page it is manipulating, showing its overall goal and options to pause or deactivate it (1). A virtual cursor represents the agent and its actions. Clicking the cursor opens a contextual menu to give instructions or fully take over control. When taken over, the virtual cursor snaps to and aligns with the user's cursor (2). Moving the cursor outside the page returns control to the agent and resumes automation.}
    \Description{An illustration shows two color-coded automation agents working on webpages in parallel. Each agent is represented by a large virtual cursor on the page and a matching colored control bar below it that states its goal. On the left, a green agent selects a date on a travel site. On the right, the user has clicked the purple agent's cursor to open a menu of commands, including an option to `Take over'. A diagram shows the agent's virtual cursor snapping to the user's actual cursor to signify that the user is now in direct control.}
    \label{fig:cursors}
\end{figure}

\rv{AI-driven retrieval (e.g., Retrieval-Augmented Generation) systems like Perplexity usually provide a single summary at the end of a session, despite supporting follow-up queries. In contrast, \mbox{\sys} enables an iterative and user-controlled process of interpreting sources. As users open and select webpages, they can surface intermediate summaries that evolve with their browsing. By toggling the \emph{\mbox{\expose}} sidebar, they can view summaries for each page or an aggregated summary across the selected set. \mbox{\sys} updates and caches these summaries whenever the selection or web content changes. These granular, dynamic summaries support interleaved sensemaking and information gathering.}

\rv{Users can refine their retrieval by asking follow-up questions about the selected pages, or the pages in view when none are selected. If the answer is not available from the current content, \mbox{\sys} deploys automation agents to navigate and search for additional context (Section~\mbox{\ref{sec:agent}}).}

\subsection{Operate at Scale: On-Demand, Contextualized, and Parallel Web Automation}
\label{sec:agent}

\begin{scenario}
\rv{After settling on her flight, hotel, and a few activities, Leela moves on to finalizing the planning. She selects all relevant pages, invokes automation agents---one per page---and asks them to complete the bookings. Each agent pulls from her user profile and the schedule and details in her notes to make the reservation in parallel. The sidebar summarization updates in real time, showing the status of each booking as it completes.}
\end{scenario}

\begin{figure}[ht]
    \centering
    \includegraphics[width=\linewidth]{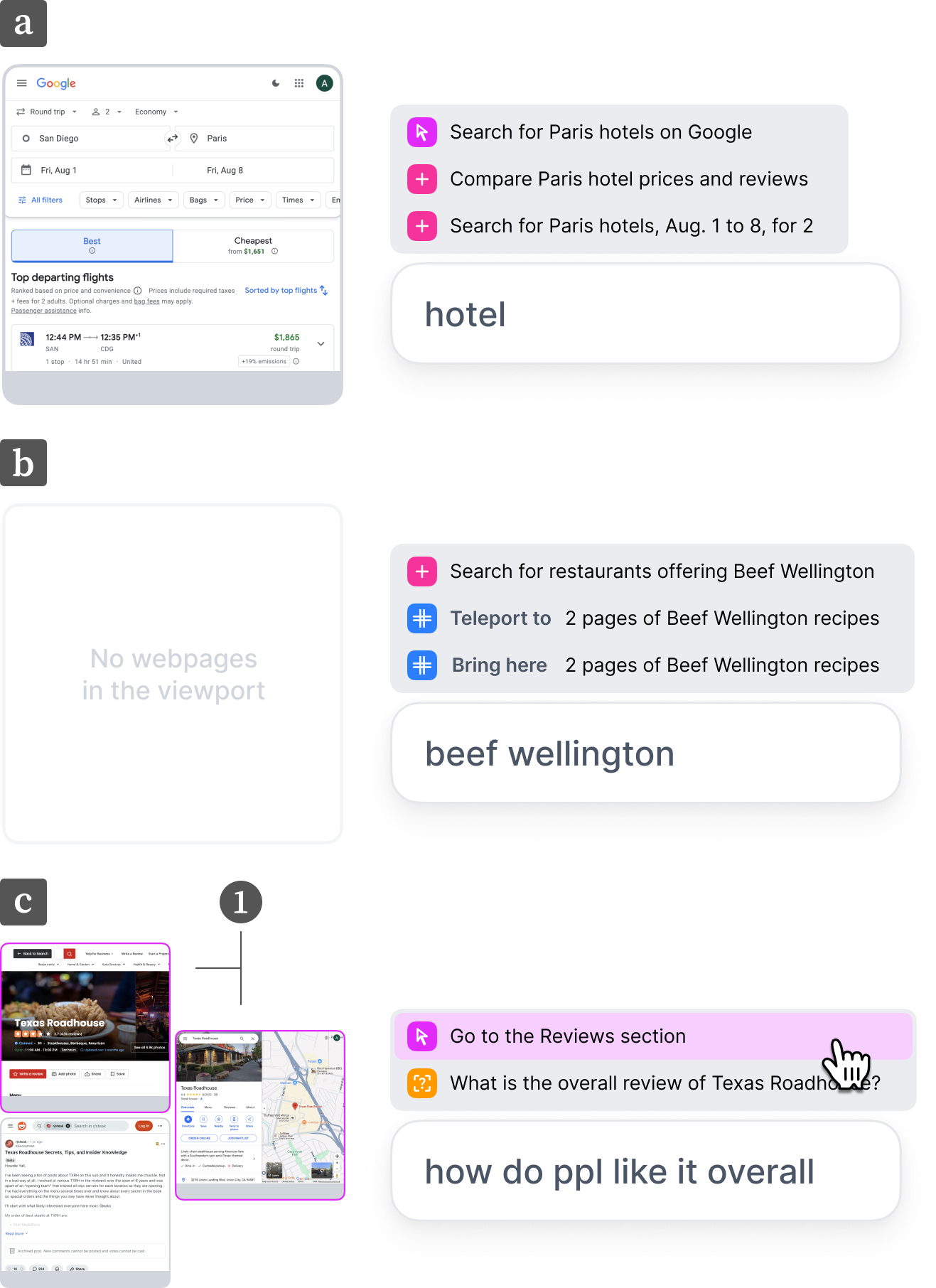}
    \caption{The global command bar with feedforward options (right) and the current webpages in the viewport (left). Options of varied types are ordered from bottom to top by relevance for quick access using the arrow keys. (a) After searching for flights to Paris, \sys uses the context to suggest relevant \emph{create} actions. (b) \emph{Organize} actions support canvas-wise content search and quick revisiting of related pages. (c) \emph{Operate} and \emph{query} actions allow the user to interact with or synthesize across pages. Hovering over an option highlights corresponding webpages on the \webcanvas (1).}
    \Description{Three examples show how a global command bar suggests actions based on the user's context. Each example displays the current webpages on the left and the command bar on the right. First, with a flight search to Paris on screen, the command bar suggests related actions like searching for hotels. Second, when the screen is empty, the command bar suggests actions to find and open new pages based on the user's query. Third, with two pages open, the command bar suggests an action to find information on one of them; hovering over the suggestion highlights the relevant page.}
    \label{fig:feedforward}
\end{figure}

\newcommand*{\mline}[1]{%
\begingroup
    \renewcommand*{\arraystretch}{1.6}%
    \begin{tabular}[c]{@{}>{\raggedright\arraybackslash}p{12.5cm}@{}}#1\end{tabular}%
\endgroup
}

\newcommand*{\mlineshort}[1]{%
\begingroup
    \renewcommand*{\arraystretch}{1.6}%
    \begin{tabular}[c]{@{}>{\centering\arraybackslash}p{3cm}@{}}#1\end{tabular}%
\endgroup
}

\begin{table*}[ht]
  \caption{Types of Feedforward Options Supported by the Global Command Bar}
  \label{tab:feedforward}
  \renewcommand{\arraystretch}{1.6} 
  \setlength{\tabcolsep}{4pt} 
  \small
  \centering
  \begin{tabular}{
    @{}>{\centering\arraybackslash}p{1.5cm}
    >{\centering\arraybackslash}p{12.5cm}
    >{\centering\arraybackslash}p{3cm}@{}
  }
    \toprule
    Types & Explanation & Corresponding Activities \\
    \midrule
  \end{tabular}

  \begin{tabular}{
    @{}>{\raggedright\arraybackslash}m{1.5cm}
    |>{\raggedright\arraybackslash}p{12.5cm}
    |>{\centering\arraybackslash}p{3cm}@{}
  }
    \raisebox{-0.2\height}{\includegraphics[height=1.1em]{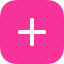}} \textbf{Create} & \mline{\textbf{Start at scale.} Generate multiple starting points in parallel to bootstrap a browsing session (such as alternative search queries or a set of suggested pages) based on the current pages in the viewport and the user's recent browsing history~\cite{xia2023crosstalk}.} & Starting \\
    \hline
    \raisebox{-0.2\height}{\includegraphics[height=1.1em]{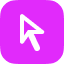}} \textbf{Operate} & \mline{\textbf{Operate at scale.} Perform task-specific actions on specific pages, such as filling in forms. When needed, multiple agents may be deployed across webpages to perform tasks simultaneously, e.g., \emph{``Navigate to the comment section for each hotel page.''}} & \mlineshort{Scouting, Filtering, Collecting, Transacting} \\
    \hline
    \raisebox{-0.2\height}{\includegraphics[height=1.1em]{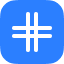}} \textbf{Organize} & \mline{\textbf{Organize at scale.} Arrange webpages in the viewport by sorting (for webpages within grid structures), selecting, closing, or locating specific webpages---either by teleporting to them or bringing them into the current viewport (Figure~\ref{fig:feedforward}.b).} & \mlineshort{Filtering, Revisiting, Collecting} \\
    \hline
    \raisebox{-0.2\height}{\includegraphics[height=1.1em]{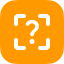}} \textbf{Query} & \mline{\textbf{Synthesize at scale.} Ask questions about the webpages in view, enabling the system to retrieve and synthesize information across them. Responses appear in the side panel, where users can continue asking follow-up questions.} & Synthesizing \\
    \bottomrule
  \end{tabular}
\end{table*}

\rv{Prior approaches to web automation typically involve agents completing full tasks end-to-end with no user involvement, such as generating a complete report for a query. However, users often struggle to stay updated on progress and cannot provide timely feedback for emerging needs and preferences~\mbox{\cite{huq2025cowpilot}}. \mbox{\sys} instead offers ad hoc, ``atomic'' automation: agents act on demand and support specific parts of a user-driven workflow.}

\rv{As described earlier, automation in \mbox{\sys} supports tasks such as navigating newly opened pages, exploring pages to gather information in response to user questions, and carrying out direct operational requests, whether globally or on selected webpages. Automation is visualized through virtual cursors---one per agent---so users can track and control actions in real time (Figure~\mbox{\ref{fig:cursors}}).}

\rv{\mbox{\webcanvas} encodes rich shared context across webpages, both visually for the user and as global context for agents. As a result, agents become automatically contextualized without requiring manual input. For example, if several webpages and a form are placed side by side, selecting them and invoking an agent to fill the form allows it to pull the context directly from the neighboring pages.}

\rv{Systems based on tabbed browsers typically rely on a single agent executing tasks sequentially, resulting in slow progress and poor traceability due to constant tab switching and disconnected views. In contrast, \mbox{\sys} leverages the \mbox{\webcanvas} as a workspace where multiple processes run and are visualized in parallel, similar to dashboards or spreadsheets (where computations occur concurrently in space~\mbox{\cite{dourish2017spreadsheets}}). For example, to check the distance to an airport for a set of hotels, all detail pages can be opened and organized into a grid, and agents can be deployed to each page to navigate to the relevant section simultaneously and extract the information.}

Since all automation runs in parallel and is shown on the canvas, users can easily monitor, intervene, or assist as needed---for example, to bypass CAPTCHAs or rescue a stuck agent. When control has shifted from the agent to the user, the virtual cursor follows the user cursor and aligns with it to indicate the change (Figure~\ref{fig:cursors}.6).

\subsection{Supporting Diverse Scalable Instructions with Feedforward Prompting}
\label{sec:feedforward}

\rv{While at-scale controls on the \mbox{\webcanvas} can be performed via direct manipulation, \mbox{\sys} also provides a global command bar for orchestrating malleable webpages using natural language (Figure~\mbox{\ref{fig:feedforward}}).}

\rv{The bar, toggled with \texttt{Command + T} or a button, behaves like a traditional address bar for navigation or search when users press \texttt{Return}.}
\rv{Building on the workflow identified in the problem framing (Section~\mbox{\ref{sec:background}}) and the requirements for browsing at scale, we derive four types of supported cross-page requests: \emph{create}, \emph{operate}, \emph{organize}, and \emph{query} (Table~\mbox{\ref{tab:feedforward}}).}

\rv{A common challenge with prompting LLM-assisted actions is their black-box interpretation~\mbox{\cite{subramonyam2024bridging,zamfirescu2023johnny,min2025feedforward}}: users often cannot predict what the model can do or how it will respond. With a defined set of cross-page actions, \mbox{\sys} supports \emph{feedforward prompting} by translating open-ended input into clear, available options for system actions. Users then select an option for AI to execute.}

During prototyping, we found that asking the LLM to suggest and rank actions across all four types in one pass was slow and often missed relevant actions in some types.
To address this, \sys sends four parallel requests, one for each type, allowing the LLM to identify actions specific to each.
\sys then aggregates the results by filtering out low-confidence actions, sorts the options, and presents the most likely ones for the user to choose from (Appendix~\ref{app:agg}).

\section{Implementation}

The \sys prototype is an Electron app built with React, where each webpage is loaded in its own \texttt{webview}~\cite{electronjswebview}. The \webcanvas interface is implemented using tldraw~\cite{tldraw}.
All LLM-based tasks, including web automation and content summarization, are powered by Claude 3.7 Sonnet (\texttt{20250219}).
For web automation, we developed custom HTML distillation and agentic pipelines, following common practices in LLM-based web automation~\cite{zhou2023webarena,he-etal-2024-webvoyager,browser2024}.
As a result, we can reuse the distilled web content for \extraction (Appendix~\ref{app:html}) and support flexible user intervention during the automated process.

\rv{Displaying many webpages can place heavy loads on the device. An M4 Max MacBook Pro with 36GB of RAM can handle about 80 webpages before freezing, which already exceeds what most users open for a typical task~\mbox{\cite{chang2021tab,ma2023browsing}}. However, when we apply techniques used in modern browsers, such as background tab pausing\footnote{https://support.google.com/chrome/answer/12929150}, which stops rendering and scripting for inactive pages while preserving the last rendered content, the system remains functional even after loading 500 tabs in testing. No participants experienced related issues during the evaluation.}

\rv{Because Electron is Chromium-based, our prototype inherits the default security model of other Chromium browsers (e.g., Chrome) for resisting malicious webpages and managing cached user data. In addition, the model providers we use state that they do not access or train on LLM requests sent through their APIs, which is how \mbox{\sys} communicates with them. Given this setup, we believe security and user privacy are reasonably protected for a research prototype, though alternatives such as locally hosted models could further strengthen privacy guarantees\footnote{Such as https://ollama.com}.}

\sys is open-sourced\footnote{https://orca.jiang.pl} to facilitate further exploration of at-scale web browsing, and to serve as a testbed for advancing parallel agentic web automation pipelines.

\section{Preliminary Evaluation}
\label{sec:evaluation}

Evaluating a new browser is challenging due to established user habits with tabbed browsing. A comprehensive understanding of many designs and features requires a longitudinal study with daily use. However, security concerns make it unsuitable for production and beyond this research's scope. \rv{To sufficiently assess the usability and effectiveness of the new design concepts and features, we conducted preliminary lab studies.}

\rv{Because conventional tabbed browsing already serves as participants' de facto baseline from daily use, we did not add a separate control condition and instead relied on their existing experience to judge the usability and perceived benefits of \mbox{\sys}.}

We aim to answer the following research questions:

\begin{enumerate}[leftmargin=2.5em, label=\bfseries RQ\arabic*]
  \item To what extent does \sys support the users in exploring, making sense of, and retrieving information on the web?
  \item How effectively can the users navigate the webpages on a spatial canvas? What challenges do they encounter?
  \item How effectively do users understand the system status, express their needs, and collaborate with AI?
  \item How is the user experience of user-driven and AI-facilitated browsing compared to AI-driven search engines?
  \item \rv{How does \mbox{\sys} help users identify and recover from AI and system failures?}
\end{enumerate}

\subsection{Participants}

We recruited 8 participants (4 female, 4 male, aged 21--27) via internal channels at a university. All participants regularly use generative AI tools like ChatGPT. \rv{They have varying experience with canvas-based graphics editing tools (e.g., Figma, Adobe Illustrator)---four participants (P1, P3, P5--6) use these tools often (more than a few times per week), P7--8 use them less than a few times per month, and P2 and P4 use them once every few months or less.}

\subsection{Procedure}

The in-person study lasted about 60 minutes per participant, with each participant receiving a 20 USD Amazon Gift Card. The study included the following phases:

\paragraph{Introduction (5 minutes)} Participants received a brief introduction to the research and an overview of \sys, covering the main interface components.

\paragraph{Semi-structured walkthrough and tutorial (30 minutes)} Participants were guided through all features with example tasks and scenarios. They interacted with the system independently with experimenter guidance, familiarizing themselves with \sys's capabilities before moving on to freeform tasks. During the walkthrough, participants were encouraged to also apply or combine features with other tasks to fully experiment with and understand \sys.

\paragraph{Freeform tasks (15 minutes)} Participants were asked to use \sys to complete a task of their choice.

\paragraph{Questionnaire and interview (10 minutes)} After the tasks, participants completed a 5-point Likert-scale questionnaire evaluating system dimensions and overall experience. A semi-structured interview gathered in-depth qualitative feedback.

\subsection{Findings}

Overall, participants found the batch operations and the overview and synthesis at scale helpful in supporting their browsing process while allowing them to stay in control.
One key finding was that when the cost of navigating, operating, and managing context across pages is reduced, users not only scout and synthesize information more quickly, but also develop a greater ``appetite'' for exploration and information foraging.
Users tend to open more tabs than before, which broadens their context and strengthens the information grounding that supports deeper understanding and better decision-making.
We summarize the detailed findings of the questionnaire, interview, and user behavior analysis below.

\subsubsection{Easier navigation and inspection at scale stimulates users' desire for exploration (RQ1)} 

All participants reported that \sys helped them explore more during information-oriented tasks (3 strongly agree and 5 agree in the Likert-scale survey).
They felt navigation and understanding across multiple pages became easier and faster. As P4 noted, \emph{``I feel more empowered browsing than with the classic browser... one click could count as many more clicks before.''} Participants are \emph{``more willing to browse more''} (P4), and with \batchopen and \expansion, they are \emph{``able to open [like this] many tabs all at once... [which was] hard to navigate [with traditional browsers], but here you can visually see what each tab is and organize it nicely''} (P3) leading them to explore more pages and options.

For example, when choosing a restaurant, P5 opened multiple Yelp lists for different cuisines side by side, used AI to open multiple options from each, and filtered and sorted the pages with \sys to find the best options for each cuisine.
\emph{``Some wanted Vietnamese food, one wanted Italian. Before, we had to decide the cuisine first, then search,''} as restarting separate searches and comparing results were time-consuming, \emph{``but now I'm definitely encouraged to search more since I'm not spending any more time and I can gather all the results and interact with them at the same time.''}

Scanning through many pages with large amounts of information was also made easier with \extraction, which participants unanimously mentioned as the most helpful feature (6 strongly agree, 2 agree).
When comparing coffee shops, P1 opened a grid of pages and added multiple abstraction queries, including rating, Wi-Fi, parking, price, hours, and outdoor seating.
\sys continuously extracted the data, allowing P1 to scan quickly without distractions: \emph{``These are all the things I think about when I look at coffee shops, and I can see them all here without scrolling up and down for each page.''}
They were then able to quickly filter out places that did not meet their needs. As a result, P1 noted, \emph{``I am no longer afraid of opening lots of pages at first, because I know I can then go through them quickly and find what I want.''}

\subsubsection{Custom use of the spatial layout helps with diverse activities of the exploration process (RQ2)}

To our surprise, all participants found the spatial layout of webpages easy to navigate and manage, regardless of their experience with graphics editing tools (5 strongly agree, 3 agree).
They appreciated that the \webcanvas provides the \emph{``bird's-eye view''} (P7) and \emph{``mind mapping''} (P8) of multiple websites, which became \emph{``notably more helpful as the number of webpages grows''} (P6), and helped them stay aware of their exploration progress and the pages they had opened (P1--3, P7).

Participants frequently referred to their experience with graphics editors.
For example, P5 described using Figma not just for design work, but to manage all project-related documents and information, where they can easily compare changes across duplicates side by side, instead of \emph{``dealing with a bunch of separate Google Docs.''}
They appreciated that now with a spatial browser and \emph{``persistent webpages,''} they could also preserve videos and interactive tech demos alongside other content in one workspace.
P1 noted that the dynamic, customizable layout of the canvas helped organize messy and divergent informational tasks.
They compared it to graphic design, pointing out that \emph{``how humans process information and workflows are never necessarily structured.''}

We observed that all participants intuitively organized webpages by task, placing them in different areas and forming distinct clusters.
P5 liked how they could \emph{``assign meaning or context to different areas''} without making concrete commitments to folders and structures.
For example, after using \sys to surface the win-loss ratios for all thirty NBA teams from their respective homepages, P3 clustered the pages (i.e. teams) into two groups---those with positive and negative ratios. P3 then used the dynamic summary feature to generate summary stats and descriptions for each cluster.

With the help of \sys's search function, they can then jump between these clusters.
For example, while filling out the questionnaire (a Google Form) in \sys, P4--8 used the global command bar to search for and jump to the webpages they wanted to reference while thinking aloud about their ratings.
Participants appreciated how spatial separation helped them maintain clean, focused workspaces without needing to close tabs. As P6 said, \emph{``I often move the tabs I don't want to work on now to a separate window... [but] if I [accidentally] close it, I lose all my tabs.''}

\subsubsection{Contextualized and atomic AI facilitation adapts to diverse preferences for human-AI collaboration (RQ3)}

Participants were convinced that \sys effectively tracks their context to provide corresponding assistance (2 strongly agree, 6 agree).
P4 was surprised to see feedforward suggestions like \emph{``Restaurants in Paris for two people''} appear after just typing \emph{``Restaurants''} in the global command bar, after searching for flights to Paris.
However, they were actually looking for local dining options for that evening.
By scrolling the viewport to move the flight search pages out of view, \sys updated the suggestions accordingly, removing Paris from the context.

On the other hand, we observed diverse preferences in how participants used AI facilitation features.
For example, P1, P2, and P7 utilized features like \batchopen and \expansion for speeding up exploration, but preferred to read pages themselves from time to time, pinning pages of interest to the side.
In contrast, P6 and P8 relied more on summaries and chatting with the AI to obtain information from pages.
These differences likely stem from the varied browsing habits people have developed over time~\cite{chang2021tab,teevan2004perfect}.
By integrating AI support into each activity of web browsing and information foraging, from scouting to synthesis, \sys adapts to users' diverse needs, allowing them to flexibly control certain aspects while delegating others to the AI.

We also observed varied attitudes toward the effectiveness of monitoring and controlling automation agents (2 strongly agree, 1 agree, 3 neutral, 2 disagree).
P2 and P7 appreciated being able to leave agents running \emph{``in the background''} (P2) or \emph{``at the periphery''} (P7) while continuing their own tasks.
P7 noted, \emph{``I feel I'm frozen looking at what [OpenAI] Operator is doing... [while] it feels opportunistic [with \sysnospace] in a way that I can do more things while waiting for the agents to complete the task,''} and added that it is easy to \emph{``keep track of multiple agents at the same time... as long as it's less than 10.''}
In contrast, P8 felt \emph{``overwhelmed when more than 3 or 4 [agents] were active.''}
P4 also found the virtual cursors distracting and preferred not to see them as long as the agents completed their tasks correctly.
The diversified responses point to the need for further investigation into user preferences and cognitive limits in monitoring and controlling real-time multi-agent systems.

\subsubsection{Direct engagement with and control over information sources fosters more personalized information retrieval and strengthens user trust in the results (RQ4)}

Participants reported feeling more directly engaged and in control of information and its sources when using \sys compared to AI-driven tools like ChatGPT and Perplexity (2 strongly agree, 6 agree).
P6 noted that as the process is more \emph{``transparent''}---they could see which links the agents opened and verify the content themselves---they \emph{``trust the results more.''}

P3 emphasized the value of being able to \emph{``do their own research''} alongside the AI, rather than \emph{``blindly''} accepting AI-generated summaries.
P1 even mentioned that they might not use the summary feature at all, since they now have access to \emph{``the actual websites.''}
Previously, they were often worried that ChatGPT might \emph{``miss certain information''} and felt \emph{``mistrustful''} of the results.

P2 appreciated that \sys allowed them to control the sources during the search. For example, when researching synthesizer options, \sys initially opened a set of relevant sites, including reviews, Reddit threads, and shopping pages.
P2 then used \expansion with only Reddit pages to gather more Reddit posts, and selected all of them to generate a summary. As P2 explained, \emph{``I know Reddit is my community, and I just want information from there,''} while reviews or shopping sites might contain sponsored content.
To compare, P2 opened an AI-driven search engine during the study. Despite being explicitly requested, it still generated a summary based on multiple mixed sources.

P4 highlighted the benefits of source-level control in tasks involving personal information and judgment, \emph{``I trust Expedia, I trust Airbnb, but not other less common ones---so even if you open a bunch, I'll only check the ones I trust.''}

\subsubsection{Continuous visibility and batch controls help users identify and recover from AI failures (RQ5)}

\rv{Because current LLM-based agents are stochastic and error-prone, failures are inevitable: they may misinterpret instructions, act on the wrong elements, or navigate to unintended pages~\mbox{\cite{huq2025cowpilot}}.}

\rv{To monitor and prune agent actions, P1 routinely range-selected webpages to be updated after each request and focused the camera on them (Figure~\mbox{\ref{fig:webpage}}.1), using the canvas as a visual checklist. They also appreciated that the camera view automatically snapped to newly created webpages, making new content and activity immediately visible. When \mbox{\expansion} produced pages that did not match expectations, P1 selected and deleted them at once. Similarly, P7 closely monitored agents and, when an agent tried to add an unwanted grocery item, deleted that agent, zoomed in on the relevant page, and completed the step manually instead.}

\rv{Meanwhile, once a page moved outside the current viewport, participants had difficulty tracking updates or agent behaviors on that page. As P7 zoomed into a single page to take over actions with a focused view, they were surprised upon zooming back out to find that many other agents had already finished their tasks elsewhere on the canvas. Although they did not observe obvious errors, this surfaced a limitation of ``off-canvas'' agents operating on pages outside the user's view. A future version could pause such off-canvas agents by default, especially for potentially dangerous actions; or add off-screen activity indicators on the minimap or screen edges, similar to cues used in digital maps.}

\rv{Participants also adapted their strategies when a feature repeatedly failed. For example, when P4 wanted to see flights from adjacent days, they first tried \mbox{\expansion} with query \emph{``find flights +-2 days''} and expected the four expanded pages to land on the correct dates. However, the model repeatedly attempted to open target pages by adding URL parameters---disallowed by Google Flights. As a remedy, P4 used global undo (\texttt{Command + Z}) to revert the operation across pages, then duplicated the original page, range-selected the copies, and commanded \emph{``navigate to days +-2 days''}. \mbox{\sys} then correctly deployed one agent per page, each navigating to a different, expected day.}

\subsubsection{Limitations}

The main limitation of our study lies in the short time participants had to learn and use \sys.
By the end of the session, participants had an average of 37 webpages on the \webcanvas.
As P5 noted, linear tabbed browsing has become ``muscle memory,'' and it can feel overwhelming to adjust to a new browsing modality within a limited time frame.
Although we did not capture any complaints about the spatial layout of webpages during the study, this may also be because the tasks were relatively contained and did not span long periods, as real-world tasks often do. Participants may not have accumulated a large enough number of pages or complex workflows to fully experience the challenges---or benefits---of navigating a dense canvas across multiple tasks.

A secure, robust, and highly performant implementation is required to further evaluate \sys's long-term usability in complex and diverse real-world settings.

\section{Discussion and Future Work}
\label{sec:discussion}

Reflecting on our design process, study findings, and limitations of the current prototype, we discuss future directions for advancing user-driven, at-scale browsing and other informational tasks.

\subsection{Blessings and Curses of Spatial Information Interfaces}
\label{sec:spatial}

\sys enables a parallel overview across webpages through a spatial canvas, while helping users access content from ``miniaturized'' pages using \extraction and dynamic summaries.
The canvas representation supports flexible, evolving workflows, allowing users to accumulate and structure content in multi-dimensional yet unconstrained ways~\cite{bederson1994pad,agarawala2006keepin}.
Meanwhile, this \emph{``freeform, organized chaos''} (P5) can lead to visual clutter. Misaligned items and overlapping elements may create extra overhead for users who want a clean, orderly workspace~\cite{beyond}. In contrast, while linear stacks are semantically messy, they are visually tidy.
Future research could explore several directions to augment spatial information interfaces:

First, improve content organization through automatic alignment.
While \sys currently supports grid and stack layouts, other structures---such as content-based categorization or localized gestures like snapping based on semantic relationships---could offer more meaningful or lightweight alignment options.

Second, enhance viewport management to reduce the manual effort of panning and zooming.
We experimented with rule-based viewport management, where the view could automatically snap to one or a set of pages.
However, it failed to deliver significant improvements, as users' focus shifted quickly during exploration, and rule-based view snapping often disrupted the fluid transitions needed for dynamic tasks.
Future work could explore content- and context-aware viewport snapping to better support fluid workflows.

Thrid, an overview across web content, or information, does not have to be a spatial one.
Future research could investigate alternative structures, such as a linear grid (similar to a spreadsheet), where users can place webpages or other content in fixed cells, to understand the tradeoffs compared to spatial layouts.

\rv{Finally, while \mbox{\sys} introduces a drastically different UI from today's browsers, it is not intended to \emph{replace} the tab-based paradigm. Instead, our exploration surfaces a complementary interface for user-driven information-foraging tasks in the era of AI. The \mbox{\webcanvas} layout can coexist with a traditional tabbed UI---for example, zooming into a page could transition to a tabbed view, whereas zooming out would return to a canvas where the page is surrounded by others. A future deployment study could examine how, given such dual modes and fluid transitions, users appropriate different views for different tasks and preferences.}

\subsection{Malleable User Interfaces for the Web}

Today's browsers primarily serve as ``renderers'' for content and interfaces predefined by website designers and developers, respecting their intended styles, priorities, and even rate limits for how content is consumed. However, these independently designed webpages often fall short in supporting diverse user needs or working seamlessly together for tasks that span across multiple sites~\cite{min2025malleable}. Moreover, interaction mechanisms built for human users are poorly suited for the growing demands of automation agents~\cite{zhou2023webarena}.

\sys treats webpages as malleable materials that users can flexibly overview, represent with different structures, and transform into succinct formats in the malleable browser space.
However, webpages are still primarily displayed as isolated ``rectangles'' that preserve the styles and formatting of the original websites, which are cluttered with irrelevant content---although \extraction offers simple information retrieval and UI component surfacing.

\rv{With rapidly advancing LLM capabilities and emerging agentic infrastructures, such as Model Context Protocol (MCP) and NLWeb, we anticipate AI agents will soon be able to navigate, operate, interpret, and extract information from most websites with greater reliability and depth~\mbox{\cite{llmstxt,mcp,nlweb,cloudflare}}. To this end, we envision treating webpages not merely as visual interfaces, but as ``services'' or APIs that provide rich data, content, and functionality.}

LLM-powered browsers can then serve as interface ``composers'' and leverage their generative capabilities to create custom, task-specific interfaces layered over the web-served content~\cite{webstrates}.
Agentic pipelines serve as a transformation layer that connects the custom UI and data of varied formats and schemas~\cite{mcp,nlweb}.
This would enable the creation of malleable, personalized, and evolving workspaces that can selectively incorporate information aligned with users' tasks across ``data sources''~\cite{cao2025jelly,min2023masonview}.

\subsection{Parallel and Browser-Level Multi-Agent Systems}
\label{sec:parallel-agents}

\rv{\mbox{\sys} introduces granular, contextualized, multi-threaded web automation by allowing multiple agents to operate in parallel across pages to complete tasks.
Because these browser-level agents can read and act on content across many tabs, they surface new privacy and security challenges, including what information they may access, how their actions are scoped across sites, and how cross-site data flows are governed at the browser level.
In our current design, agents are invoked via explicit user commands on selected page subsets as their permitted context, and their actions remain visible and interruptible in real time, providing containment, oversight, and user control.}

On the other hand, participant feedback revealed a need for better monitoring and control of real-time multi-agent systems.
P6 suggested a global progress bar to track all agents, rather than scattered across the \webcanvas.
P2 proposed breaking down and visualizing the agents' detailed steps using an interactive diagram to make it easier for users to track and adjust their behavior~\cite{jiang2023graphologue,majeed2024steering}.

\rv{Therefore, future research can explore new mechanisms to monitor agent progress, manage permissions, and enable timely user intervention. This could involve methods such as direct manipulation, demonstrations, or even a ``group chat'' interface for users to give intermediate feedback to one or more agents simultaneously, complemented by workspace-level permission models and UI that explicitly scope which pages, data, and operations each agent is allowed to access~\mbox{\cite{lau2009programming}}. We see fine-grained visibility, controllability, and scoping as central to making browser-level multi-agent orchestration not only more usable, but also more privacy-preserving and secure, critical for it to scale and be adopted by millions.}

\section{Conclusion}
\label{sec:conclusion}

We introduce \sys, a prototype web browser that reconceptualizes webpages as malleable materials and supports user-driven, AI-facilitated orchestration across them in the malleable browser space, aiding web-based informational work.

With a spatial canvas interface, \sys provides features to enable browsing at scale across webpages, addressing the distributed nature of web-based informational tasks.
Our evaluation shows that \sys reduces the effort required to manage complex browsing workflows while preserving user agency, increasing engagement, and stimulating users' desire for information foraging. These findings highlight a promising direction for future browser designs that tightly integrate AI capabilities into the personal, user-driven informational tasks.

\begin{acks}
We thank members of the Foundation Interface Lab for their support and generous help during the production of the work.
\end{acks}

\balance
\bibliographystyle{ACM-Reference-Format}
\bibliography{main}

\newpage

\appendix

\nobalance

\section{Constructing Feedforward Option Lists from Parallel LLM Responses}
\label{app:agg}

To reduce a user prompt to concrete actions the system can perform, \sys sends
four parallel prompts, one for each category (create, organize, operate, and query),
to the LLM. The model returns candidate actions with confidence scores, and the system
aggregates them into the feedforward options list as follows.

Let $C=\{c_{1},c_{2},c_{3},c_{4}\}$ be the set of categories. For every $c_{i}$
the LLM yields
\[
  A_{i}=\{\, (a_{ij},\,s_{ij}) \mid s_{ij}\in[0,1]\,\}.
\]

\paragraph{Filtering}
Discard low-confidence items and merge the rest:
\[
  A=\bigcup_{i=1}^{4}\bigl\{\, (a_{ij},s_{ij})\in A_{i}\mid s_{ij}\ge0.2 \bigr\}
  .
\]

\paragraph{Sorting}
Sort the remaining in descending order of the scores:
\[
  A_{\text{sorted}}=\bigl[(a_{1},s_{1}),\dots,(a_{|A|},s_{|A|})\bigr].
\]

\paragraph{Selecting the final list}
Index the list with $k$. Define the running total of confidence scores of selected
actions as
\[
  S_{n}=\sum_{k=1}^{n}s_{k},
\]
and choose the smallest prefix length
\[
  N=\min\bigl\{\,n\in\{1,\dots,|A_{\text{sorted}}|\}\,\big|\, S_{n}\ge3\text{ or
  }n=7\bigr\}.
\]
The final action set is
\[
  A^{*}= \{\, (a_{k},s_{k})=A_{\text{sorted}}[k] \mid 1\le k\le N \}.
\]

\section{HTML Distillation for Web Automation and Page Extraction}
\label{app:html}

Following common practices in HTML distillation, we develop a custom
distillation algorithm to convert the often highly complex HTML structures of
production websites into a concise and manageable form. This distilled representation
allows the text-based \sys agents to reason over and act on web content effectively.
We incorporate several notable approaches, briefly described below.

We annotate interactive elements with two interaction types: \code{click} (such
as buttons) and \code{input} (for all kinds of \code{<input />} elements, such as
text fields or checkboxes). We also include the \code{aria-label} or \code{placeholder}
text when available and applicable. Each interactive element is annotated using
the following syntax:
\providecommand{\ul}[1]{\underline{#1}}
\providecommand{\setul}[2]{}
\begin{center}
  \texttt{[\$\ul{id}\$:\ul{interactionType} (\ul{ariaLabel})] \ul{content}}
\end{center}

For example:
\begin{center}
  \texttt{[\$m2l\$:click (Send email)] Send}
\end{center}

The \code{id} is a stable, short, and unique identifier derived from the
element's \code{xpath}, ensuring consistent reference across distillations.

We further annotate active elements---typically a button that has just been
clicked or an input box that is currently focused---as:
\begin{center}
  \texttt{[ACTIVE] >{>}> [\$m2l\$:click (Send email)] Send <{<}<}
\end{center}

Changes in elements since the last distillation are shown in a separate prompt
following the full current HTML, using a code diff style, such as:
\begin{center}
  \begin{tabular}{@{}l@{}}
    \texttt{[-] [\$m2l\$:click (Send email)] Send} \\
    \texttt{[+] [\$6jp\$:click] Cancel Send} \\
  \end{tabular}
\end{center}

This helps the LLM focus on parts of the interface that are likely to be most relevant,
such as a newly appeared pop-up modal.

\paragraph{Referring to elements for automation agents}
Agents can refer to elements using their stable identifiers when executing automation
steps. For example:
\begin{center}
  \begin{tabular}{@{}l@{}}
    \texttt{\{ "action": "click",}           \\
    \texttt{\ \ "element": "nuu" \}}         \\[1em]
    \texttt{\{ "action": "update-value",}    \\
    \texttt{\ \ "element": "617",}           \\
    \texttt{\ \ "value": "Hello, World!" \}}
  \end{tabular}
\end{center}

\paragraph{Referring to elements for \extraction}
Similarly, when an interactive element is extracted from the original page, represented
as a standardized UI element, the underlying ``target'' element is referred to
using the identifier as well. For example:
\begin{center}
  \begin{tabular}{@{}l@{}}
    \texttt{\{ "type": "button",}   \\
    \texttt{\ \ "title": "Send",}   \\
    \texttt{\ \ "target": "m2l" \}}
  \end{tabular}
\end{center}

That is, a button titled ``Send'' will be extracted, and clicking it will
propagate the click event to the underlying element \code{m2l}.

\end{document}